\documentclass[12pt]{article}
\usepackage{amsfonts}
\usepackage{amssymb}
\usepackage{graphics,psboxit,amsmath}
\usepackage{subfigure}
\usepackage{graphicx}
\usepackage{verbatim}

\def\hybrid{\topmargin -30pt    \oddsidemargin 0pt 
        \headheight 0pt \headsep 0pt
        \textwidth 6.25in       
        \textheight 9.5in       
        \marginparwidth .875in
        \parskip 5pt plus 1pt   \jot = 1.5ex}

\hybrid

\def\baselinestretch{1.2}

\catcode`\@=11

\def\marginnote#1{}
\newcount\hour
\newcount\minute
\newtoks\amorpm
\hour=\time\divide\hour by60
\minute=\time{\multiply\hour by60 \global\advance\minute by-\hour}
\edef\standardtime{{\ifnum\hour<12 \global\amorpm={am}%
        \else\global\amorpm={pm}\advance\hour by-12 \fi
        \ifnum\hour=0 \hour=12 \fi
        \number\hour:\ifnum\minute<10 0\fi\number\minute\the\amorpm}}
\edef\militarytime{\number\hour:\ifnum\minute<10 0\fi\number\minute}

\def\draftlabel#1{{\@bsphack\if@filesw {\let\thepage\relax
   \xdef\@gtempa{\write\@auxout{\string
      \newlabel{#1}{{\@currentlabel}{\thepage}}}}}\@gtempa
   \if@nobreak \ifvmode\nobreak\fi\fi\fi\@esphack}
        \gdef\@eqnlabel{#1}}
\def\@eqnlabel{}
\def\@vacuum{}
\def\draftmarginnote#1{\marginpar{\raggedright\scriptsize\tt#1}}

\def\draft{\oddsidemargin -.5truein
        \def\@oddfoot{\sl preliminary draft \hfil
        \rm\thepage\hfil\sl\today\quad\militarytime}
        \let\@evenfoot\@oddfoot \overfullrule 3pt
        \let\label=\draftlabel
        \let\marginnote=\draftmarginnote
   \def\@eqnnum{(\theequation)\rlap{\kern\marginparsep\tt\@eqnlabel}%
\global\let\@eqnlabel\@vacuum}  }

\def\draft2{
        \def\@oddfoot{\sl preliminary draft \hfil
        \rm\thepage\hfil\sl\today\quad\militarytime}
        \let\@evenfoot\@oddfoot \overfullrule 3pt
        \let\label=\draftlabel
        \let\marginnote=\draftmarginnote
   \def\@eqnnum{(\theequation)\rlap{\kern\marginparsep\tt\@eqnlabel}%
\global\let\@eqnlabel\@vacuum}  }

\def\preprint{\twocolumn\sloppy\flushbottom\parindent 2em
        \leftmargini 2em\leftmarginv .5em\leftmarginvi .5em
        \oddsidemargin -.5in    \evensidemargin -.5in
        \columnsep .4in \footheight 0pt
        \textwidth 10.in        \topmargin  -.4in
        \headheight 12pt \topskip .4in
        \textheight 6.9in \footskip 0pt
        \def\@oddhead{\thepage\hfil\addtocounter{page}{1}\thepage}
        \let\@evenhead\@oddhead \def\@oddfoot{} \def\@evenfoot{} }

\def\numberbysection{\@addtoreset{equation}{section}
        \def\theequation{\thesection.\arabic{equation}}}

\def\underline#1{\relax\ifmmode\@@underline#1\else
        $\@@underline{\hbox{#1}}$\relax\fi}

\def\titlepage{\@restonecolfalse\if@twocolumn\@restonecoltrue\onecolumn
     \else \newpage \fi \thispagestyle{empty}\c@page\z@
        \def\thefootnote{\fnsymbol{footnote}} }

\def\endtitlepage{\if@restonecol\twocolumn \else \newpage \fi
        \def\thefootnote{\arabic{footnote}}
        \setcounter{footnote}{0}}  

\catcode`@=12
\relax

\def\figcap{\section*{Figure Captions\markboth
        {FIGURECAPTIONS}{FIGURECAPTIONS}}\list
        {Figure \arabic{enumi}:\hfill}{\settowidth\labelwidth{Figure
999:}
        \leftmargin\labelwidth
        \advance\leftmargin\labelsep\usecounter{enumi}}}
 \relax
\def\tablecap{\section*{Table Captions\markboth
        {TABLECAPTIONS}{TABLECAPTIONS}}\list
        {Table \arabic{enumi}:\hfill}{\settowidth\labelwidth{Table
999:}
        \leftmargin\labelwidth
        \advance\leftmargin\labelsep\usecounter{enumi}}}
 \relax
\def\reflist{\section*{References\markboth
        {REFLIST}{REFLIST}}\list
        {[\arabic{enumi}]\hfill}{\settowidth\labelwidth{[999]}
        \leftmargin\labelwidth
        \advance\leftmargin\labelsep\usecounter{enumi}}}
 \relax
\makeatletter
\newcounter{pubctr}
\def\publist{\@ifnextchar[{\@publist}{\@@publist}}
\def\@publist[#1]{\list
        {[\arabic{pubctr}]\hfill}{\settowidth\labelwidth{[999]}
        \leftmargin\labelwidth
        \advance\leftmargin\labelsep
        \@nmbrlisttrue\def\@listctr{pubctr}
        \setcounter{pubctr}{#1}\addtocounter{pubctr}{-1}}}
\def\@@publist{\list
        {[\arabic{pubctr}]\hfill}{\settowidth\labelwidth{[999]}
        \leftmargin\labelwidth
        \advance\leftmargin\labelsep
        \@nmbrlisttrue\def\@listctr{pubctr}}}
 \relax
\makeatother

\def\be{\begin{equation}}
\def\ee{\end{equation}}
\def\ba{\begin{eqnarray}}
\def\ea{\end{eqnarray}}

\def\del{\partial}

\def\r{\rho}
\def\a{\alpha}

\def\b{\beta}

\def\g{\gamma}
\def\G{\Gamma}
\def\d{\delta}

\def\P{\Pi}

\def\th{\theta}
\def\Th{\Theta}
\def\m{\mu}

\def\Om{\Omega}

\def\s{\sigma}
\def\S{\Sigma}

\def\cN{{\cal N}}

\def\elK{{\bf K}}
\def\elPi{{\bf \Pi}}
\def\elE{{\bf E}}

\def\no{\noindent}

\def\qq{\qquad}

\def\IR{\relax{\rm I\kern-.18em R}}

\def \ha {{1\over 2}}

\def \ov {\over}

\def\const{{\rm const.}}

\begin{document}


\renewcommand{\theequation}{\thesection.\arabic{equation}}
\csname @addtoreset\endcsname{equation}{section}

\newcommand{\eqn}[1]{(\ref{#1})}
\begin{titlepage}
\begin{center}

\hfill hep--th/0612139\\

\vskip .5in

{\Large \bf Stability of strings dual to flux tubes\\between
static quarks in $\cN=4$ SYM}

\vskip 0.5in

{\bf Spyros D. Avramis$^{1,2}$},\phantom{x} {\bf Konstadinos
Sfetsos}$^1$\phantom{x} and\phantom{x} {\bf Konstadinos Siampos}$^1$
\vskip 0.1in

${}^1\!$
Department of Engineering Sciences, University of Patras,\\
26110 Patras, Greece\\

\vskip .1in

${}^2\!$
Department of Physics, National Technical University of Athens,\\
15773, Athens, Greece\\

\vskip .15in

{\footnotesize {\tt avramis@mail.cern.ch}, \ \ {\tt sfetsos@upatras.gr},
\ \ {\tt ksiampos@upatras.gr}}\\

\end{center}

\vskip .4in

\centerline{\bf Abstract} \no Computing heavy quark-antiquark
potentials within the AdS/CFT correspondence often leads to
behaviors that differ from what one expects on general physical
grounds and field-theory considerations. To isolate the
configurations of physical interest, it is of utmost importance to
examine the stability of the string solutions dual to the flux
tubes between the quark and antiquark. Here, we formulate and
prove several general statements concerning the perturbative
stability of such string solutions, relevant for static
quark-antiquark pairs in a general class of backgrounds, and we
apply the results to ${\cal N}=4$ SYM at finite temperature and at
generic points of the Coulomb branch. In all cases, the
problematic regions are found to be unstable and hence physically
irrelevant.

\vfill
\no


\end{titlepage}
\vfill
\eject


\tableofcontents

\def\baselinestretch{1.2}
\baselineskip 20 pt
\no

\section{Introduction}

In the framework of the AdS/CFT correspondence \cite{adscft}, the
computation of the static Wilson-loop quark-antiquark potential in
$\cN=4$ super Yang--Mills at large 't Hooft coupling is mapped to
the classical problem of minimizing the action for a string
connecting the quark and antiquark on the boundary of ${\rm
AdS}_5$ and extending into the radial direction. This approach was
first employed in \cite{maldaloop} for the conformal case and
further extended in \cite{wilsonloopTemp,bs} to non-conformal
cases. Moreover, motivated by the recent interest in applying
AdS/CFT ideas and techniques to calculations relevant to moving
mesons in thermal plasmas, the finite-temperature calculations of
this type have been extended in various ways
\cite{PSZ}--\cite{AEV}.

\no Although in the conformal case these calculations reproduce
the expected Coulomb behavior of the potential, in the
non-conformal extensions there are cases where the behavior of the
potential is quite different from what one anticipates based on
expectations from the field-theory side. For instance, for
finite-temperature $\cN=4$ SYM, one encounters multiple branches
of the quark-antiquark potential as well as a behavior resembling
a phase transition, while for the Coulomb branch of $\cN=4$ SYM
one encounters a linear confining behavior at certain regions of
the moduli space. To check whether these types of behavior are
actually predicted by the gauge/gravity correspondence, one must
subject the corresponding string configurations to certain
consistency checks, one of which is their stability under small
fluctuations.

\no For the conformal case, the equations of motion for the
fluctuations have been first obtained in \cite{cg,kmt} and result
in a positive spectrum as expected. For the finite-temperature
case, where two branches of the solution occur, the question of
stability was investigated recently in \cite{michalogiorgakis} for
the case of a {\em moving} quark-antiquark pair. For this problem,
the small-fluctuation analysis yielded a system of coupled
differential equations, and numerical investigations suggested
that the long string corresponding to the upper branch of the
potential is unstable against small perturbations in the
longitudinal directions.

\no Here we will conduct a stability analysis, restricting
ourselves to static configurations and pursuing an analytic
treatment all the way. We will investigate perturbations in all
coordinates including the angular ones and we will consider
$\cN=4$ SYM both at finite temperature and at the Coulomb branch.
In this setting, the equations for the longitudinal, transverse
and angular perturbations decouple and the problem is amenable to
analytic methods. In particular, the strategy that we will employ
is to write the equations of motion satisfied by the fluctuations
in both Sturm--Liouville and Schr\"odinger forms and to deduce the
frequency spectrum of the fluctuations either from the
Sturm--Liouville zero modes or from the Schr\"odinger potentials,
using both exact and approximate methods.

\no This article is organized as follows. In section 2, we
describe the calculation of Wilson loops in the AdS/CFT framework
via classical string configurations in D3-brane backgrounds and we
discuss the general features of the resulting quark-antiquark
potentials. In section 3, we consider small fluctuations about the
classical solutions, we turn the equations of motion of these
fluctuations into one-dimensional Schr\"odinger problems, and we
establish several general results for the exact and approximate
calculation of the corresponding energy eigenvalues. In section 4,
we specialize the general calculations of section 2 to the cases
of non-extremal and multicenter D3-branes and we identify certain
regions of the parameter space for which the potentials fail to
capture the behavior expected on physical grounds. In section 5,
we apply the general stability analysis to these configurations
and we prove that all regions where problematic behavior arises
are perturbatively unstable. In section 6, we summarize and
conclude. Finally, in the appendix we discuss an interesting
analogy of our problem with the standard classical-mechanical
problem of determining the shapes of a soap film stretching
between two rings and their stability.

\section{The classical solutions}
\label{sec-2}

Our general setup refers to the AdS/CFT calculation of the static
potential of a heavy quark-antiquark pair according to the
well-known recipe of \cite{maldaloop} for the conformal case and
its extensions \cite{wilsonloopTemp,bs} beyond conformality. On
the gauge-theory side, this potential is extracted from the
expectation value of a rectangular Wilson loop with one temporal
and one spatial side. On the gravity side, the Wilson loop
expectation value is calculated by extremizing the Nambu--Goto
action for a fundamental U-shaped string propagating into the dual
supergravity background, whose endpoints are constrained to lie on
the two temporal sides of the Wilson loop. Below, we give a brief
review of this procedure, following \cite{bs}, and we outline the
main qualitative features of the resulting potentials.

\no
We consider a general metric of the form
\be
\label{2-1} ds^2 = G_{tt} dt^2 + G_{yy} dy^2 + G_{uu} du^2 +
G_{xx} dx^2 + G_{\th\th} d\th^2 + \ldots\ ,
\ee
noting that we will use Lorentzian signature throughout the paper.
Here, $y$ denotes the (cyclic) coordinate along which the spatial
side of the Wilson loop extends, $u$ denotes the radial direction
playing the r\^ole of an energy scale in the dual gauge theory and
extending from the UV at $u \to \infty$ down to the IR at some
minimum value $u_{\rm min}$ determined by the geometry, $x$ stands
for a generic cyclic coordinate, $\th$ stands for a generic
coordinate on which the metric components may depend in a
particular way to be specified shortly and the omitted terms
involve coordinates that fall into one of the two latter classes.
Note that the metric \eqn{2-1} is diagonal and that we do not
consider off-diagonal terms in the present paper. For the analysis
that follows, it is convenient to introduce the functions
\be
\label{2-2}
g(u,\th) = - G_{tt} G_{uu}\ ,\qq f_y(u,\th) = - G_{tt} G_{yy}\ ,
\ee
while for the stability analysis, we also introduce
\be
\label{2-3}
f_x(u,\th) = - G_{tt} G_{xx}\ ,\qq f_\th(u,\th) = - G_{tt} G_{\th\th}\ ,
\qq h(u,\th) = G_{yy} G_{uu}\ .
\ee
It is useful to mention the behavior of the above functions in the
conformal limit, where the metric reduces to that of ${\rm AdS}_5
\times {\rm S}^5$. Using the leading order expressions
\be
-G_{tt} \simeq G_{xx} \simeq G_{yy} \simeq {u^2 \ov R^2} \ ,\quad G_{uu}
\simeq {R^2\ov u^{2}} \ , \quad G_{\th\th} \simeq R^2\ ,\qq {\rm as} \quad
u\to \infty\ , \label{ba0}
\ee
we see that
\be
g \simeq h \simeq 1\ ,\quad f_x \simeq f_y \simeq u^4\ ,\quad
f_\th \simeq u^2 \ , \qq {\rm as} \quad u\to \infty\ . \label{gh1}
\ee

\no
In the framework of the AdS/CFT correspondence, the interaction
potential energy of the quark-antiquark pair is given by
\be
\label{2-4}
e^{-{\rm i} E T} = \langle W(C) \rangle = e^{{\rm i} S[C]}\ ,
\ee
where
\be
\label{2-5}
S[C] = - {1 \ov 2 \pi} \int d \tau d \sigma \sqrt{- \det g_{\a \b} }\ ,
\qq g_{\a\b} = G_{\mu\nu}
\partial_\alpha x^\mu \partial_\b x^\nu \ ,
\ee
is the Nambu--Goto action for a string propagating in the dual
supergravity background whose endpoints trace the contour $C$. To
proceed, we fix reparametrization invariance by choosing
\be
\label{2-6}
t=\tau \ ,\qq u=\s \ ,
\ee
we assume translational invariance along $t$, and we consider the
embedding
\be
\label{2-7}
y = y(u)\ ,\qq x =0\ ,\qq \th = \th_0 = \const\ ,\qq \hbox{rest} = \const\ ,
\ee
supplemented by the boundary condition
\be
\label{2-8}
u \left(\pm {L \ov 2} \right) = \infty\ ,
\ee
appropriate for a quark placed at $y=-L/2$ and an antiquark placed
at $y=L/2$. In the ansatz \eqn{2-7}, the constant value $\th_0$ of
the non-cyclic coordinate $\th$ must be consistent with the
corresponding equation of motion. As we shall see later on, this
requires that
\be
\label{2-9}
\partial_\th g(u,\th) |_{\th=\th_0} = \partial_\th f_y(u,\th) |_{\th=\th_0} = 0\ ,
\ee
which is definitely satisfied if the metric components obey identical to the above vanishing
relations. We note that there are occasions where
the latter stronger condition is not satisfied but \eqn{2-9} is \cite{AS}.
For the ansatz given above, the Nambu--Goto action reads
\be
\label{2-10}
S = - {T \ov 2 \pi} \int d u \sqrt{ g(u) + f_y(u) y^{\prime 2}}\ ,
\ee
where $T$ denotes the temporal extent of the Wilson loop, the
prime denotes a derivative with respect to $u$ while $g(u) \equiv
g(u,\th_0)$ and $f_y(u) \equiv f_y(u,\th_0)$ are the functions in
\eqn{2-2} evaluated at the chosen constant value $\th_0$ of $\th$.
To seek a classical solution $y_{\rm cl}(u)$, we proceed by
reducing the problem to quadrature. Independence of the Lagrangian
from $y$ implies that the associated momentum is conserved,
leading to the equation
\be
\label{2-11} {f_y y_{\rm cl}^\prime \ov \sqrt{ g + f_y y_{\rm
cl}^{\prime 2}}} = \pm f_{y0}^{1/2}\qq \Longrightarrow \qq y_{\rm
cl}^\prime = \pm {\sqrt{f_{y0} F}\ov f_y}\ ,
\ee
where $u_0$ is the value of $u$ at the turning point, $f_{y0}
\equiv f_y(u_0)$, $y_{\rm cl}$ is the classical solution with the
two signs corresponding to the two symmetric branches around the
turning point, and $F$ stands for the shorthand
\be
\label{2-14} F = {g f_y \ov f_y - f_{y0}}\ .
\ee
Integrating \eqn{2-11}, we express the separation length as
\be
\label{2-12}
L = 2 f_{y0}^{1/2} \int_{u_0}^{\infty} d u {\sqrt{F} \ov f_y}\ .
\ee
Finally, inserting the solution for $y_{\rm cl}^\prime$ into
\eqn{2-10} and subtracting the divergent self-energy contribution
of disconnected worldsheets, we write the interaction energy as
\be
\label{2-13}
E = {1 \ov \pi} \int_{u_0}^\infty d u \sqrt{F} - {1 \ov \pi}
\int_{u_{\rm min}}^\infty d u \sqrt{g}\ .
\ee
Ideally, one would like to evaluate the integrals \eqn{2-12}
and \eqn{2-13} exactly, solve \eqn{2-12} for $u_0$ and insert
into \eqn{2-13} to obtain an expression for the energy $E$ in
 terms of the separation length $L$. However, in practice this
cannot be done exactly, except for the simplest possible cases,
 and Eqs. \eqn{2-12} and \eqn{2-13} are to be regarded as
parametric equations for $L$ and $E$ with parameter $u_0$.
The resulting quark-antiquark potential $E(L)$ is
required to satisfy the concavity condition \cite{concavity}
\be
\label{2-15}
{d E \ov d L} > 0\ ,\qq {d^2 E \ov d L^2} \leqslant 0\ ,
\ee
which is valid for any gauge theory, irrespective of its gauge
group and matter content, and signifies that the force is always
attractive and its magnitude is a never increasing function of the
separation distance. In our case it has been shown that \cite{bs}
\be
{dE\ov dL}= {1 \ov 2 \pi} f_{y0}^{1/2} > 0 \ ,\qq {d^2E\ov dL^2}=
{1\ov 4\pi} {f^\prime_{y0} \ov f^{1/2}_{y0}} {1 \ov
L^\prime(u_0)}\ ,
\ee
and, since in all known examples we have $f^\prime_{y0} > 0$, the
concavity condition restricts the physical range of the
integration constant $u_0$ to the values where the inequality
\be
L^\prime(u_0) < 0\
\ee
is satisfied.

\no In the archetypal example of \cite{maldaloop} for $\cN=4$ SYM,
dual to a stack of D3-branes, Eq. \eqn{2-12} can be inverted for
$u_0$ yielding a single-valued, monotonously decreasing, function
of $L$ whose substitution into \eqn{2-13} gives the expected
Coulomb potential, {\em albeit} with a coefficient which is
proportional to $\sqrt{g_{\rm YM}^2 N}$ rather than $g_{\rm YM}^2
N$, presumably as a result of strong-coupling physics. This
Coulombic behavior persists in the UV limit for all cases
involving supergravity solutions that become asymptotically ${\rm
AdS}_5\times {\rm S}^5$ as $u\to \infty$. However, towards the IR,
the behavior of the potential is quite different and depends on
the details of the supergravity solution. In particular, for
$\cN=4$ SYM at finite temperature, dual to a stack of non-extremal
D3-branes \cite{wilsonloopTemp}, as well as for the Coulomb branch
of $\cN=4$ SYM, dual to multicenter distributions of D3-branes
\cite{bs}, one encounters three types of behavior, described
below.
\begin{itemize}
\item A solution for $u_0=u_0(L)$ exists only for $L$ below a
maximal value $L_{\rm c}=L(u_{0\rm c})$ and is a double-valued
function of $L$. This signifies the existence of {\em two}
classical string configurations satisfying the same boundary
conditions, called a ``short'' string for $u_0>u_{0\rm c}$ and a
``long'' string for $u_0<u_{0\rm c}$, and satisfying
$L^\prime(u_0)<0$ and $L^\prime(u_0)>0$ respectively. Accordingly,
$E(L)$ is a double-valued function with the lower and upper
branches corresponding to the short and long strings respectively.
Although the upper branch is energetically disfavored, the
classical analysis does not clarify whether the long string is
perturbatively unstable or just metastable and hence physically
realizable. The latter possibility would have been physically
disturbing since it is conflict with the concavity condition
\eqn{2-15} due to the fact that $L^\prime(u_0)>0$ for this branch.
Such a behavior has been encountered for $\cN =4$ SYM at finite
temperature and at certain regions of the Coulomb branch.

\item A solution for $u_0$ exists only for $L$ below a maximal
value $L_{\rm c}$ and it is a single-valued function of $L$.
$E(L)$ is a single-valued function describing a screened Coulombic
potential. However, the screening length is heavily dependent on
the orientation of the string with respect to the brane
distribution.

\item  A solution for $u_0$ exists for all $L$ and $E(L)$ is a
single-valued function of $L$ interpolating between a Coulombic and a linear
confining potential. However, a confining behavior is
unexpected for the dual gauge theories under
consideration due to the underlying conformal structure in combination
with maximal supersymmetry.

\no The last two types of behavior have been encountered for some
regions of the Coulomb branch of $\cN=4$ SYM. Note that in these
cases it is not the concavity condition \eqn{2-15} that has been
violated. The discrepancies arise due to the facts that $\cN=4$ SYM at zero
temperature is not expected to be a confining theory and that the screening length should not be a
concept that depends heavily on the particular trajectory we use
to compute the heavy quark interaction potential.
\end{itemize}
In all three cases, the behavior of the interaction energy as a
function of the separation is quite puzzling and calls for a
careful interpretation. For the first case under consideration, it
is obvious that a small-fluctuation analysis will settle the issue
whether the long string solution is perturbatively unstable. One
might then be tempted to repeat the analysis for the two other
cases, where there is no {\it a priori} indication about the
stability of the solutions on energetic grounds. As it will turn
out, such an analysis will cast as unstable the parametric regions
that give rise to a heavily orientation-dependent screening length
(in the second case), and a linear behavior (in the third case).
It is quite impressive, in our opinion, that the small-fluctuation
analysis suffices to resolve the puzzles in all three cases.

\section{Stability analysis}
\label{sec-3}

Having described the basic features of the classical string
solutions of interest, we now turn to a stability analysis of
these configurations, with the aim of isolating the regions that
are of physical interest and for which the gauge/gravity
correspondence can be trusted. In this section we give a general
description of the small-fluctuation analysis for the
configurations discussed above with metric of the form \eqn{2-1}
and we establish a series of results which will ultimately allow
us to identify the stable and unstable regions by a combination of
exact and approximate analytic methods.

\subsection{Small fluctuations}

To investigate the stability of the string configurations of
interest, we consider small fluctuations about the classical
solutions discussed above. In particular, we will be interested in
three types of fluctuations, namely (i) ``transverse''
fluctuations, referring to cyclic coordinates transverse to the
quark-antiquark axis such as $x$, (ii) ``longitudinal''
fluctuations, referring to the cyclic coordinate $y$ along the
quark-antiquark axis, and (iii) ``angular'' fluctuations,
referring to the special non-cyclic coordinate $\th$. To
parametrize the fluctuations about the equilibrium configuration,
we perturb the embedding according to
\be
\label{3-1}
x = \d x (t,u)\ ,\qq y = y_{\rm cl}(u) + \d y (t,u)\ ,\qq \th = \th_0 + \d \th(t,u)\ .
\ee
keeping the gauge choice \eqn{2-6} unperturbed by using worldsheet
reparametrization invariance.\footnote{Alternatively, we could
perturb the choice of gauge by setting $u = \s + \d u(t,\s)$,
keeping $y$ fixed to $y_{\rm cl}(\s)$ as in
\cite{michalogiorgakis}. The advantage of our choice is that the
differential equation for $\d y$ is simpler than that for $\d u$.
However, as we shall see in subsection 3.2, in determining the
boundary conditions the discussion is facilitated by using $\d u$
instead of $\d y$. For details on the various possible gauges, see
\cite{kinar}.} We then calculate the Nambu--Goto action for this
ansatz and we expand it in powers of the fluctuations. The
resulting expansion
 is written as
\be
\label{3-2}
S = S_0 + S_1 + S_2 + \ldots\ ,
\ee
where the subscripts in the various terms correspond to the
respective powers of the fluctuations. The zeroth-order term gives
just the classical action. The first-order contribution is easily
seen to be
\be
\label{3-3} S_1 = - {1 \ov 2\pi} \int dt du \left[ \sqrt{f_{y0}}\
\d y^\prime + \left( {1 \ov 2 F^{1/2}} \partial_\th g + {f_{y0}
F^{1/2}
 \ov 2 f_y^2} \partial_\th f_y \right) \d \th \right]\ ,
\ee
where $\partial_\th g$ and $\partial_\th f_y$ stand for the
$\th$--derivatives of the full functions $g(u,\th)$ and
$f_y(u,\th)$, evaluated at $\th=\th_0$. The first term is a
surface contribution which may cancelled out by adding to the
classical action the boundary term which will not affect the
classical equations of motion. In the second term, the coefficient
of $\d \th$ is just the equation of motion of $\th$ and the
requirement that it vanish leads indeed to the condition
\eqn{2-9}. Using this condition and performing straightforward
manipulations, we write the second-order contribution as
\ba
\label{3-4} \!\!\!\!\!\!\!\!\!\!\!\! S_2 &=& - {1 \ov 2\pi} \int
dt du \biggl[ {f_x \ov 2 F^{1/2}}
\d x^{\prime 2} - {h f_x F^{1/2} \ov 2 g f_y} \d \dot{x}^2 \nonumber\\
 && \qq\qq\quad\; + {g f_y \ov 2 F^{3/2}} \d y^{\prime 2} - {h \ov 2 F^{1/2}} \d \dot{y}^2 \\
&& \qq\qq\quad\; + {f_\th \ov 2 F^{1/2}} \d \th^{\prime 2} - {h
f_\th F^{1/2} \ov 2 g f_y} \d \dot{\th}^2 + \left( {1 \ov 4
F^{1/2}}
\partial_\th^2 g
+ {f_{y0} F^{1/2} \ov 4 f_y^2} \partial_\th^2 f_y \right) \d \th^2
\biggr]\ ,\nonumber
\ea
where all functions and their $\th$--derivatives are again
evaluated at $\th=\th_0$. We observe that, by virtue of \eqn{2-9},
the
 various types of fluctuations decouple, which greatly facilitates
the analysis. Writing down the equations of motion for this action,
 using independence of the various functions from $t$, and
 introducing an $e^{-{\rm i} \omega t}$ time dependence by setting
\be
\label{3-5}
\d x^\m (t,u) = \d x^\m (u) e^{-{\rm i} \omega t}\ ,
\ee
we obtain the following linearized equations for the three types
of fluctuations
\ba
\label{3-6}
&&\left[ {d \ov du} \left({f_x \ov F^{1/2}} {d \ov du} \right)
+ \omega^2 {h f_x F^{1/2} \ov g f_y} \right] \d x = 0\ ,
\nonumber\\
&&\left[ {d \ov du} \left( {g f_y \ov F^{3/2}} {d \ov du} \right)
+ \omega^2 {h \ov F^{1/2}} \right] \d y = 0\ ,
\\
&&\left[ {d \ov du} \left({f_\th \ov F^{1/2}} {d \ov du} \right)
 + \left( \omega^2 {h f_\th F^{1/2} \ov g f_y}
- {1 \ov 2 F^{1/2}} \partial_\th^2 g - {f_{y0} F^{1/2} \ov 2
f_y^2} \partial_\th^2 f_y \right) \right] \d \th = 0\ . \nonumber
\ea
Therefore, the problem of determining the stability of the string
configurations of interest has reduced to a standard eigenvalue
problem for the differential operators referring to the three
types of fluctuation. More precisely, we are dealing with
differential equations of the general Sturm--Liouville type
\be
\label{3-7} \left\{ - {d \ov du} \left[ p(u;u_0) {d \ov du}
\right] - r(u;u_0)
 \right\} \Phi(u) = \omega^2 q(u;u_0) \Phi(u)\ ,\quad u_{\rm min}\leqslant u_0\leqslant u
< \infty\ ,
\ee
where the functions $p(u;u_0)$, $q(u;u_0)$ and $r(u;u_0)$ are read
off from \eqn{3-6} and depend parametrically on $u_0$ through the
function $F$ in \eqn{2-14}. Our problem then is to determine the
range of values of $u_0$ for which $\omega^2$ is negative,
signifying an instability of the classical solution.

\no Although in many cases the Sturm--Liouville description given
above is sufficient for our purposes, in other cases it will be
convenient to transform our problem into a Schr\"odinger one. To
do so, we employ the changes of variables
\be
\label{3-8} x = \int_u^\infty du \sqrt{q\ov p}\ = \int_u^\infty
du \sqrt{h\ov f_y-f_{y0}} \ ,\qq \Phi = (p q)^{-1/4} \Psi\ ,
\ee
where we note that the expression for $x$ is valid for all three
types of fluctuations. Then, Eq. \eqn{3-7} is transformed to a
standard Schr\"odinger equation
\be
\label{3-9} \left[ -{d^2 \ov dx^2} + V(x;u_0) \right] \Psi(x) =
\omega^2 \Psi(x)\ ,
\ee
with the potential
\be
\label{3-10}
V = -{r \ov q} + {p^{1/4} \ov q^{3/4}} {d \ov du}
\left[ \left( {p \ov q} \right)^{1/2} {d \ov du}
 (p q)^{1/4} \right] =  -{r \ov q} + (p q)^{-1/4} {d^2 \ov dx^2} (p q)^{1/4}\ ,
\ee
expressed as a function $V(u;u_0)$ of $u$ in the first relation
and as a function $V(x;u_0)$ of $x$ in the second one. The
Schr\"odinger problem is defined in the range $x \in [0,x_0]$
where $x_0$ is given by
\be
\label{hj1}
x_0 = \int_{u_0}^\infty du \sqrt{q\ov p} \ =
\int_{u_0}^\infty du \sqrt{h \ov f_y-f_{y0}} \ ,
\ee
and is finite, as can be verified using the asymptotic expressions
\eqn{gh1} and the fact that $p/q \sim u-u_0$ as $u\to u_0$. The
precise behavior of $x_0$ for finite values of the parameter $u_0$
depends on the details of the underlying supergravity solution.
However, using the asymptotic expressions \eqn{gh1} we deduce that
\be
\label{uvx0} x(u_0)\simeq {\G(1/4)^2\ov 4 \sqrt{2 \pi} \ u_0} \ ,
\qq {\rm as}\quad u_0\to \infty\ .
\ee
The fact that the size of the interval is finite implies that the
fluctuation spectrum is quantized and the fact that it becomes
narrower in the UV implies that there cannot be any instability in
the conformal limit of the theory. With the transformation
\eqn{3-8} the UV and IR regions are mapped to the regions near
$x=0$ and near $x=x_0$ respectively. In the Schr\"odinger
description, our problem is to determine the range of $u_0$ for
which the ground state of the potential in \eqn{3-10} has negative
energy. Note that, in general, the first of \eqn{3-8} does not
lead to a closed expression for $u$ in terms of $x$ and therefore
it is not always possible to write down the potential as an
explicit function of $x$. Nevertheless, in many cases the
expression for the potential as a function of $u$ provides useful
information by itself (for example, if $V(u;u_0)$ is manifestly
positive then the energy eigenvalues $\omega^2$ are always
positive and no instability occurs), while its transcription in
terms of $x$ may be easily accomplished when performing
perturbation-theory calculations.

\subsection{Boundary conditions}

To fully specify our eigenvalue problem, we must impose
appropriate boundary conditions on the fluctuations at the UV
limit $u\to\infty$ ($x=0$) and at the IR limit $u=u_0$ ($x=x_0$).
Starting from the Sturm-Liouville description, the boundary
condition at the UV is particularly easy to determine by looking
at the singularity structure of the differential equations in
\eqn{3-6} as $u\to \infty$. Indeed, at this limit we find
\be
\Phi^{\prime\prime} + {4\ov u} \Phi^\prime = {\cal O}(u^{-4}) \
,\qq {\rm as} \quad u\to \infty\ ,\label{3-14}
\ee
which implies that $u=\infty$ is a regular singular point and that
the two independent solutions of any equation in \eqn{3-6} have
the form
\ba
\Phi_1 &= & \sum_{n=0} c_n u^{-n}\ ,
\nonumber\\
\Phi_2 & = & d \ln u + {1\ov u^3} \sum_{n=0}^\infty d_n u^{-n}\ .
\ea
For the first solution, which is regular at infinity, we use the
freedom to assign to the constant $c_0$ any value we want to set
it to zero. For the second solution, which blows up at infinity,
we choose $d=0$. Hence we impose the boundary condition
\be
\Phi(u) = 0\ ,\qq {\rm as}\quad u\to \infty \ . \label{3-14b}
\ee
The nature of the boundary condition at the IR needs special
attention due to the fact that we have to glue the fluctuations
around the upper and lower branches of the classical solution
$y_{\rm cl}$, corresponding to the two different signs in
\eqn{2-11}. The point $u=u_0$ is again a regular singular point
and the two independent solutions admit expansions of the
form\footnote{We assume that $u_0> u_{\rm min}$ strictly. In case
of equality the singularity structure is different and will be
examined in general and in the examples of particular interest
below.}
\ba
\Phi_{1} & = & (u-u_0)^{\r} \left[ c_0 + \sum_{n=1}^\infty c_n (u-u_0)^n\right]\ ,
\nonumber\\
\Phi_{2}  & = & d_0 + \sum_{n=1}^\infty d_n (u-u_0)^n\ ,
\label{fhdh}
\ea
where $\r = 1/2 $ for the fluctuations along $x$ and $\th$ and
$\r=-1/2$ for those along $y$. We demand that, on the two sides of
the classical solution, the fluctuations and their first
derivatives with respect to the classical solution be equal.
However, this should be done not using the classical solution
$y_{\rm cl}$, but that combined with its perturbation as in
\eqn{3-1}. Equivalently, defining a new coordinate $\bar u$ as
\be
u=\bar u + \d u(t,u) \ ,\qq \d u(t,u) = - {\d y(t,u)\ov
y^\prime_{\rm cl}(u)}\ ,
\ee
the classical solution is not perturbed at all. This redefinition
does not affect the $x$-- and $\th$--fluctuations since they have
trivial classical support and we keep only linear in fluctuations
terms. However, since $y_{\rm cl}^\prime \sim (u-u_0)^{-1/2}$ near
$u=u_0$, the fluctuation $\d u$ has an expansion as in \eqn{fhdh},
but with $\r =1/2$. Since, around the matching point, the two
branches of the classical solution differ by a sign while the
fluctuations are equal, we have
\be
{d\Phi\ov d y_{\rm cl}}\bigg |_{u=u_0} = 0\ ,
\ee
where $\Phi$ refers to the $\d x,\d\th$ and $\d u$ fluctuations.
Recasting everything in terms of the original fluctuations we see
that, in the expansions \eqn{fhdh}, the coefficient $c_0=0$ for
the $x$ and $\th$ fluctuations and $d_0=0$ for the $y$ ones.
Equivalently,
\ba
\label{3-18} \d x,\d \th  : & & \phantom{xxx}
(u-u_0)^{1/2}\Phi^\prime = 0 \  ,\qq {\rm as}\quad u\to u_0^+ ,
\nonumber\\
\d y  : & & \phantom{xxx} \Phi + 2 (u-u_0) \Phi^\prime = 0 \ ,\qq
{\rm as}\quad u\to u_0^+ \ .
\ea

\no These boundary conditions for the Sturm--Liouville function
$\Phi$ can be transcribed to boundary conditions for the
Schr\"odinger wavefunction $\Psi$ by using the asymptotic
relations $u\sim 1/x$ for $u\to \infty$ and $u-u_0 \sim (x_0-x)^2$
for $u \to u_0$ and $x \to x_0$, which follow from \eqn{3-8} and
the asymptotic behavior of the $p$-- and $q$--functions. After
some manipulations, we find that for all types of fluctuations we
must have
\be
\label{3-19} \Psi(0) = 0 \ ,\qq \Psi^\prime(x_0)= 0 \ ,
\ee
which correspond to a Dirichlet and a Neumann boundary condition,
in the UV and IR respectively.

\subsection{Zero modes}

The method described above would in principle allow us to
determine the regions of stability of the solutions, provided that
the relevant Sturm--Liouville or Schr\"odinger problems could be
solved exactly. However, in many cases, these problems are quite
complicated and the spectrum is impossible to determine. On the
other hand, it turns out that we can obtain useful information by
studying a simpler problem, namely the zero-mode problem of the
associated differential operators. In what follows, we prove that
(i) transverse zero modes do not exist, (ii) longitudinal zero
modes are in one-to-one correspondence with the critical points of
the function $L(u_0)$, and (iii) the angular zero-mode spectrum
can be obtained to good accuracy by approximating the
corresponding Schr\"odinger potential by an infinite square well.
These results are crucial for the stability analysis of section 5.

\subsubsection{Transverse zero modes}

We consider first the case of the transverse fluctuations. The zero mode
solution obeying \eqn{3-14b} is up to a multiplicative constant
\ba
\d x &  = & \int^{\infty}_u {du\ov f_x} \sqrt{gf_y\ov f_y-f_{y0}}
\nonumber\\
& = & -2 {\sqrt{g f_y}\ov f_x f_y^\prime} \sqrt{f_y-f_{y0}} - 2
\int_u^\infty du \sqrt{f_y-f_{y0}} \del_u \left(\sqrt{g f_y}\ov
f_x f_y^\prime \right)
\\
& = &  - 2 \int_{u_0}^\infty du \sqrt{f_y-f_{y0}} \ \del_u
\left(\sqrt{g f_y}\ov f_x f_y^\prime \right)
  - {2\ov f_{x0}}\ \sqrt{g_0 f_{y0}\ov f^\prime_{y0}} (u-u_0)^{1/2} + {\cal O}(u-u_0)\ ,
\nonumber
\ea
where in the second step we performed a partial integration. The
zero mode solution exists provided that we can satisfy the
boundary condition \eqn{3-18}, i.e. make the coefficient of
$(u-u_0)^{1/2}$ vanish for some value of the parameter $u_0$. It
is easy to see that this is not possible and so transverse zero
modes do not exist. Therefore, if the lowest eigenvalue of the
Schr\"odinger operator corresponding to the transverse
fluctuations is positive for some value of $u_0$, it will stay
positive throughout. Since, as it will turn out, this is indeed
the case for all situations under consideration, the classical
solutions are stable under transverse perturbations.

\subsubsection{Longitudinal zero modes}

We next turn to the longitudinal fluctuations, for which we will
prove the powerful result that normalizable zero modes exist only
at the values of $u_0$ where the length function $L(u_0)$ in
\eqn{2-12} has a critical point, i.e. $L^\prime(u_0) = 0$. The
importance of this result lies in the fact that it implies that a
point where the lowest eigenvalue $\omega^2$ changes sign must
necessarily be a critical point of the length function.

\no To establish the connection between longitudinal zero modes
and critical points of the length function, we begin by writing
the longitudinal zero-mode solution obeying \eqn{3-14}. Up to a
multiplicative constant, we have
\ba
\d y &  = & \int^{\infty}_u du {\sqrt{gf_y}\ov (f_y-f_{y0})^{3/2}}
\nonumber\\ & = & 2 {1\ov f^\prime_y } \sqrt{g f_y\ov f_y-f_{y0}}
+ 2 \int_u^\infty {du\ov \sqrt{f_y-f_{y0}}} \del_u \left(\sqrt{g
f_y}\ov f_y^\prime \right)
\label{3-28}\\
& = &   2 \sqrt{g_0 f_{y0}\ov f_{y0}^{\prime 3}}\ (u-u_0)^{-1/2}
  +   2 \int_{u_0}^\infty { du\ov \sqrt{f_y-f_{y0}}}
\ \del_u \left(\sqrt{g f_y}\ov f_y^\prime \right)
 + {\cal O}\left((u-u_0)^{1/2}\right)\ .
\nonumber
\ea
Then, the boundary condition \eqn{3-18} implies that this mode
exists only if
\be
\label{3-30} \int_{u_0}^\infty { du\ov \sqrt{f_y-f_{y0}}} \ \del_u
\left(\sqrt{g f_y}\ov f_y^\prime \right) = 0  \ .
\ee
We next differentiate the length function \eqn{2-12} with respect
to $u_0$. The result is easily seen to be
\ba
\label{3-11} L^\prime(u_0) & = & {f_{y0}^\prime \ov \sqrt{f_{y0}}}
\int_{u_0}^\infty du {\sqrt{gf_y}\ov (f_y-f_{y0})^{3/2}}
 - 2 f_{y0}^{1/2} {F^{1/2} \ov f_y} \biggr|_{u = u_0}
\nonumber\\
&=&  2 {f_{y0}^\prime \ov \sqrt{f_{y0}}} \int_{u_0}^\infty { du\ov
\sqrt{f_y-f_{y0}}} \ \del_u \left(\sqrt{g f_y}\ov f_y^\prime
\right)
\\
&& +
{2 \ov f_{y0}^{1/2}} \lim_{u \to u_0} \left[ F^{1/2}
\left( {\partial_{u_0} f_{y0} \ov \partial_u f_y}
- {f_{y0} \ov f_y} \right) \right]\ ,
\nonumber
\ea
where as before we have performed a partial integration.
The last line is zero, so that
\be
L^\prime (u_0) =  2 {f_{y0}^\prime \ov \sqrt{f_{y0}}}
\int_{u_0}^\infty { du\ov \sqrt{f_y-f_{y0}}} \ \del_u
\left(\sqrt{g f_y}\ov f_y^\prime \right)\ , \label{3-32}
\ee
which is manifestly finite, in contrast with the first line in
\eqn{3-11} where both terms diverge. Comparing \eqn{3-30} and
\eqn{3-32} we see that at the extrema of $L(u_0)$ there is a zero
mode, as advertised. As a result, given the critical points of the
length function, we can determine the sign of the lowest
eigenvalue $\omega^2$ in all regions by expanding the
Schr\"odinger potential in $u_0$ about each critical point $u_{0
\rm c}$ and determining whether $\omega^2$ changes sign there.

\no There is an alternative way to understand the occurrence of
the critical value $u_{0 \rm c}$ in the above analysis. It is not
difficult to prove that the Schr\"odinger potential has the
following behavior at the extreme values of $u$, namely that
\be
\label{3-31} V(u;u_0)  =   2 u^2 \ , \qq {\rm as} \quad u\to
\infty\ ,
\ee
and
\be
\label{3-33} V_0(u_0)=V(u_0;u_0)  =  {f_{y0}^{\prime\prime}\ov 2
h_0} + {h_0^\prime f_{y0}^\prime \ov 8 h_0^2} -{3\ov 8}
{f_{y0}^\prime \ov h_0}\left({g_0^\prime \ov g_0}+{f_{y0}^\prime
\ov f_{y0}} \right) \ .
\ee
Since the potential rises from a minimum value to infinity in the
finite interval $x\in [0,x_0]$, where $x_0$ is given by \eqn{hj1},
the spectrum of fluctuations is discrete. In addition, if its
minimum value $V_0$ is negative enough the potential can very well
support bound states with negative energy and this happens
whenever \eqn{3-30} has a solution for some values of the
parameter $u_0$. The largest of these values is $u_{0 \rm c}$ in
which the lowest energy eigenvalue becomes zero.

\subsubsection{Angular zero modes}

We finally consider the angular zero modes which, unlike the two
other types of zero modes, cannot be explicitly written in an
integral form due to the presence of the ``restoring force'' term
in the corresponding Sturm--Liouville equation at the third line of
\eqn{3-6}. On the other hand, in the Schr\"odinger description, we
can obtain the full fluctuation spectrum to quite high accuracy by
approximating the corresponding potential by an infinite well. The
procedure is described below.

\no We first note that the behavior of the angular Schr\"odinger
potentials at the limits $u \to \infty$ and $u=u_0$ in all our
examples is as follows
\be
V_\infty (u_0) \equiv V(\infty;u_0) = \ha \lim_{u\to \infty}
u^2\del_\th^2 g\  ,
\ee
and
\be
V_0(u_0) \equiv V(u_0;u_0)  = {1\ov 8} {g_0 f_{y0} f_{y0}^\prime
\ov h_0^2 f_{\th 0}^2}\ \del_{u_0} \left(h_0 f_{\th 0}^2\ov g_0
f_{y0}\right) + \ha {g_0\ov h_0 f_{\th 0}} \del^2_\th f_{y0}\ .
\ee
where the limit in the first equation is finite. Since the
potential rises from a minimum value to infinity in the finite
interval $x\in [0,x_0]$, where $x_0$ is given by \eqn{hj1}, we may
approximate it by an infinite well given by
\be
\label{kwsft1} V_{\rm approx} =
\left\{ \begin{array}{l l} \ha (V_0+V_{\infty}) , \quad 0\leqslant x \leqslant x_0\\
\infty\  ,\quad {\rm otherwise} \end{array}\right\}\ .
\ee
With the boundary conditions \eqn{3-19} the energy spectrum is given by
\be
\label{3-34} \omega^2_n(u_0) = {\pi^2 (2 n + 1)^2\ov 4
[x_0(u_0)]^2} + \ha \left[V_0(u_0)+V_{\infty}(u_0)\right] \ ,\qq n=0,1,\dots \ .
\ee
Note that we have taken the average value of the values of the
potential at the two extreme values of $u$. In order for a
solution to exist, the above average must be negative at least in
a finite range of values for $u_0$. The critical value $u_{0 \rm
c}$ is then obtained by numerically solving the equation
$\omega^2_0(u_0)=0$. We note that different choices for $V_{\rm
approx}$ are possible (e.g. the average value of the potential
itself) but, conceptually, they do not offer something new.

\no The advantage of the method just described is that it gives
approximately the entire spectrum of fluctuations and its
dependence on the parameter $u_0$ without the need for more
sophisticated numerics. In particular, we may plot the lowest
eigenvalue $\omega^2_0$ as a function of $u_0$ and easily check
whether it is monotonously increasing function, as it is in all of
our examples. In addition we will see that, in one case, the
infinite-well approximation is in fact exact. Finally, we will
show that there is a hierarchy of different critical values $u_{0
{\rm c},n}$, $n=0,1,\dots$, with $u_{0 \rm c} \equiv u_{0 {\rm
c},0}$ and $u_{0 {\rm c},n}>u_{0 {\rm c},n+1}$, at which the
corresponding eigenvalues $\omega^2_n$ become zero and below which
they become negative.

\no At this point one might wonder whether this infinite-well
approximation could be applied to the longitudinal fluctuations,
using, for example, \eqn{3-33} for the value of the potential at the
bottom of the well, to yield in addition the full approximate
spectrum of fluctuations. For our specific examples, it can be
shown that, although such a choice gives a reasonable
approximation to the result for $u_{0 \rm c}$ following from
\eqn{3-30}, it also results in a tower of negative eigenvalues as
we lower $u_0$, which is in conflict with our general conclusions
following from \eqn{3-33}. The discrepancy is traced to the fact
that the true potential varies significantly during the interval
$[0,x_0]$ and thus supports fewer negative-energy states than the
infinite-well potential used to approximate it. One may of course
devise phenomenological potentials that mimic the expected
behavior, but there is no need to do so in the present context.

\subsection{Perturbation theory}

As stated earlier on, the stability of the solutions against
longitudinal perturbations will be determined by identifying the
critical points of $L(u_0)$ and solving the Schr\"odinger equation
for small deviations of $u_0$ from each critical point $u_{0 \rm
c}$ by means of perturbation theory. In doing so, we must note
that the parameter $u_0$ enters into the problem not only by
appearing explicitly in the potential $V(x;u_0)$, but also by
controlling the size of the interval $x_0$ in which the problem is
defined. In what follows, we will present the perturbation-theory
formulas that are appropriate for such a case.

\no To this end, we consider a more general Schr\"odinger problem
defined on the interval $[0,x_0]$ with a potential that depends on
a parameter $u_0$, for which we know that for some $u_0=u_{0 \rm
c}$ and $x_0=x_{0 \rm c}$ the Schr\"odinger equation \eqn{3-9}
admits a single solution with a given eigenvalue $\omega^2$ (equal
to zero in our cases of interest). We would like to determine the
correction to the energy eigenvalue when $u_0$ and $x_0$ deviate
from $u_{0 \rm c}$ and $x_{0 \rm c}$. An easy computation shows
that the potential can be written as
\ba
V(x;u_0) & = & V(x;u_{0 \rm c})+ \d V(x)\ ,
\nonumber\\
 \d V(x)& = & {\d x_0 \ov x_{0 \rm c}}\Big[2 V(x;u_{0 \rm c})+ x \del_x V(x;u_{0 \rm c})\Big]
+ \d u_0 \del_{u_{0 \rm c}}V(x;u_{0 \rm c}) + \dots \ ,
\label{3-21}
\ea
where $x\in[0,x_{0 \rm c}]$ while $\d u_0 = u_0 - u_{0 \rm c}$ and
$\d x_0 = x_0 - x_{0 \rm c}$. Then, a careful computation, keeping
track of boundary terms, gives the energy shift
\ba
\label{gp} \d \omega^2 & = & \int_{0}^{x_{0 \rm c}} dx\
|\Psi(x)|^2 \d V(x)
\nonumber\\
& = & {\d x_0 \ov x_{0 \rm c}}\ \left\{ 2 \omega^2 + \left[ \ha
(\Psi^* \Psi^\prime + \Psi \Psi^{*\prime}) -x\Big[ |\Psi^\prime|^2
+ (\omega^2-V) |\Psi|^2\Big]\right]_{0}^{x_{0 \rm c}} \right\}
\nonumber\\
&& +\ \d u_0\ \int_{0}^{x_{0 \rm c}} dx\ |\Psi(x)|^2\ \del_{u_{0
\rm c}} V(x;u_{0 \rm c})\ ,
\ea
where the term $(\omega^2-V) |\Psi|^2$ can be alternatively
written as $-\ha \left(\Psi^{\prime\prime} \Psi^* +
\Psi^{*\prime\prime} \Psi\right)$ using the Schr\"odinger
equation. In deriving this, we have to use the boundary conditions
so that the unperturbed Hamiltonian is Hermitian, which is always
the case when the boundary conditions are Dirichlet, Neumann or a
linear combination thereof.

\no
To apply this result to our case, we set $\omega^2=0$ and we
note that the parameters $x_0$ and $u_0$ are related by \eqn{hj1}.
Then, the energy shift is found to be
\be
\d \omega^2 = \d u_0 \left[ x^\prime_0(u_{0 \rm c}) V(x_{0 \rm
c};u_{0 \rm c})|\Psi(x_{0 \rm c})|^2 + \int_{0}^{x_{0 \rm c}} dx\
|\Psi(x)|^2\ \del_{u_{0 \rm c}} V(x;u_{0 \rm c}) \right]\ ,
\label{3-22}
\ee
with $x_{0 \rm c}=x_0(u_{0 \rm c})$ and where we have used the boundary
conditions \eqn{3-19}. Regarding the first term, an
explicit computation, using \eqn{hj1}, gives
\be
x^\prime_0(u_0) = f^\prime_{y0} \int_{u_0}^\infty du\
{\del_u\left(h^{1/2} {f^\prime_y}^{-1}\right) \ov
\sqrt{f_y-f_{y0}}}\ . \label{j1}
\ee
Turning to the second term we must note that the $u_0$--derivative
acts on the potential $V(x;u_0)$ whose explicit form is not known.
To transcribe this expression into one involving the potential
$V(u;u_0)$, for which explicit expressions are available, we must
also take into account that when varying $u_0$ while keeping $x$
constant, $u$ varies as well. That is to say, $\del_{u_0}
V(x;u_0)$ is actually the ``convective'' derivative
\be
\del_{u_0} V(x;u_0) = \del_{u_0} V(u;u_0) + {\del u\ov \del
{u_0}}\ \del_u V(u;u_0)\ ,
\ee
where the second term can be evaluated with the aid of the
expression
\be
{\del u\ov \del u_0}= {f^\prime_{y0}\ov f^\prime_y} +
f^\prime_{y0} \sqrt{f_y-f_{y0}\ov h} \int_{u}^\infty du\
{\del_u\left(h^{1/2} {f^\prime_y}^{-1}\right) \ov
\sqrt{f_y-f_{y0}}}\ . \label{j2}
\ee
Note that, in the case where \eqn{3-22} gives $\d \omega^2=0$, we
have to go beyond first-order perturbation theory to determine
whether the energy eigenvalue changes sign. Luckily, no such
behavior occurs in our examples.

\section{D3-brane backgrounds: The classical solutions}
\label{sec-4}

In this section, we review the behavior of the quark-antiquark
potentials emerging in Wilson-loop calculations for non-extremal
 and multicenter D3-brane backgrounds. The purpose of this review
is to identify the three types of problematic behavior referred to
at the end of section 2, which motivated our stability analysis.
Since we will work with the Nambu--Goto action, we need mention in
the expressions below only the metric and not the self-dual
five-form which is the only other non-trivial field present in our
backgrounds.

\subsection{Non-extremal D3-branes}

We start by considering a background describing a stack of $N$
non-extremal D3-branes. The field-theory limit of the metric reads
\be
\label{4-1} ds^2 = {u^2 \ov R^2} \left[ - \left( 1 - {\m^4 \ov
u^4} \right)
 dt^2 + d \vec{x}_3^2 \right]
+ R^2 \left( {u^2 \ov u^4 - \m^4}\ du^2 + d \Omega_5^2 \right)\ ,
\ee
where the horizon is located at $u=\m$ and the Hawking temperature
is $T={\m \ov \pi R^2}$. This metric is just the direct product of
${\rm AdS}_5$--Schwarzschild with ${\rm S}^5$ and it is dual to
 $\cN=4$ SYM at finite temperature. For the calculations that
follow, it is convenient to switch to dimensionless variables by
rescaling all quantities using the parameter $\m$. Setting $u \to
\m u$ and $u_0 \to \m u_0$ and introducing dimensionless length
and energy parameters by
\be
\label{4-2} L \to {R^2 \ov \m} L\ ,\qq E \to {\m \ov \pi} E\ ,
\ee
we see that all dependence on $\m$ and $R$ drops out so that
 we may set $\m \to 1$ and $R \to 1$ in what follows. The
functions in \eqn{2-2} and \eqn{2-3} depend only on $u$
(reflecting the fact that all values of $\th$ are equivalent) and
are given by
\be
\label{4-3}
g(u) = 1\ ,\qq f_y(u) = u^4-1\ ,
\ee
and
\be
\label{4-4}
f_x(u) = u^4-1\ ,\qq f_\th(u) = {u^4-1 \ov u^2}\ ,
\qq h(u) = {u^4 \ov u^4 -1}\ ,
\ee
respectively. From this it also follows that the equations of
motion \eqn{2-9} for $\th$ are identically satisfied for all
values of $\th_0$.

\no Let us now evaluate the quark-antiquark potential according to
the guidelines of section 2. The integrals for the length and
energy are given by \cite{wilsonloopTemp,bs}
\begin{figure}[!t]
\begin{center}
\begin{tabular}{cc}
\includegraphics[height=5.2cm]{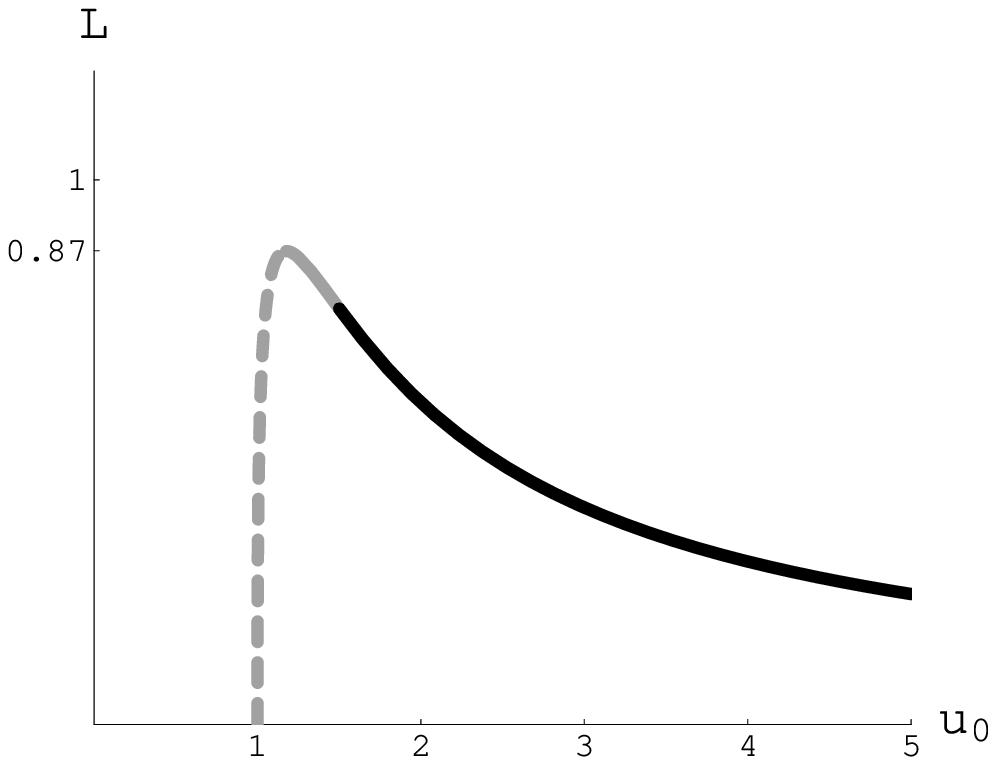}
&\includegraphics[height=5.2cm]{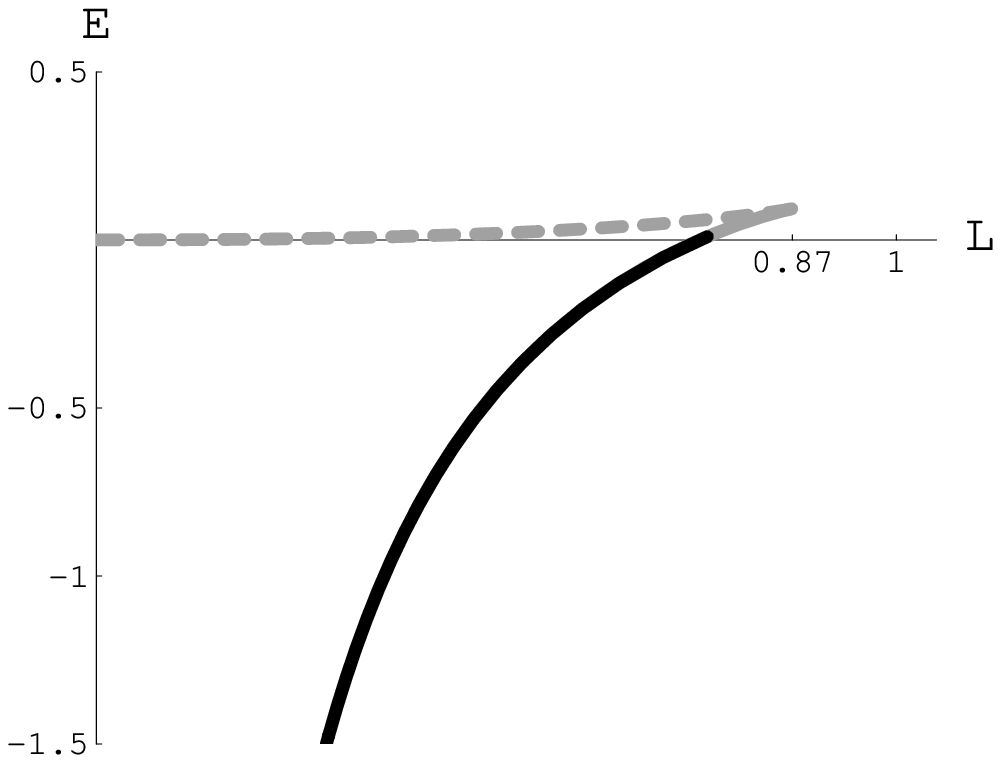}\\
(a) & (b)
\end{tabular}
\end{center}
\vskip -.5 cm \caption{Plots of $L(u_0)$
 and $E(L)$ for the non-extremal D3-brane. In this and subsequent
 plots, the various types of lines correspond to stable (solid dark),
 metastable (solid gray) and unstable (dashed gray) configurations,
 with the stability determined by the analysis of section 5.}
\label{fig1}
\end{figure}
\be
\label{4-5}
L = 2\sqrt{u_0^4-1} \int_{u_0}^\infty {du \ov
 \sqrt{(u^4-1)(u^4-u_0^4)}} = {2 \sqrt{2} \pi^{3/2}
\ov \G(1/4)^2}{\sqrt{u_0^4-1} \ov u_0^3}
 {}_2F_1\left( \ha,{3 \ov 4},{5 \ov 4};{1 \ov u_0^4} \right)\ ,
\ee
and
\be
\label{4-6} E = \int_{u_0}^\infty du \left( \sqrt{{u^4-1 \ov
u^4-u_0^4}} -1 \right)  - (u_0-1)= - {\sqrt{2} \pi^{3/2} \ov
\G(1/4)^2} u_0\,  {}_2F_1\left( - \ha,- {1 \ov 4},{1 \ov 4};{1 \ov
u_0^4} \right) + 1\ ,
\ee
where ${}_2F_1(a,b,c;x)$ is the hypergeometric function and $u_0
\geqslant 1$. For $u_0 \gg 1$ the behavior is Coulombic while at
the opposite limit, $u_0 \to 1$, we have the asymptotics
\be
\label{4-7} L \simeq \sqrt{u_0-1} \left( \ln {8 \ov u_0-1} -
 {\pi \ov 2} \right)\ ,\qq E \simeq {u_0-1 \ov 2}
\left( \ln {8 \ov u_0-1} - 1 - {\pi \ov 2} \right)\ .
\ee
The function $L(u_0)$ has a single global maximum, which together
with the corresponding maximal length and energy are given by
\cite{bs}
\be
 u_{\rm c} \simeq 1.177\ , \qq  L_{\rm c} \simeq  0.869\ ,\qq
E_c \simeq 0.093\ . \label{4-8a}
\ee
For $L > L_{\rm c}$, only the disconnected solution exists. For $L
< L_{\rm c}$, Eq. \eqn{4-5} has two solutions for $u_0$,
corresponding to a short and a long string respectively and,
accordingly, $E$ is a double-valued function of $L$. Moreover,
there exists another value of the length, given in our case by
$\tilde{L}_{\rm c} \simeq 0.754$, above which the disconnected
configuration becomes energetically favored and the short string
becomes metastable, which can be used as an alternative definition
of the screening length. The behavior described above is shown in
the plots of Fig. \ref{fig1} and, provided that the upper branch
of $E(L)$ is physically irrelevant, corresponds to a screened
Coulomb potential.

\subsection{Multicenter D3-branes}

We now proceed to the case of multicenter D3-brane distributions.
These were first constructed as the extremal limits of rotating
D3-brane solutions \cite{cy,rs} in \cite{trivedi,sfet1} and belong
to the rich class of continuous distributions of M- and string
theory branes on higher dimensional ellipsoids \cite{Basfe2}.
These distributions have been used in several investigations
within the AdS/CFT correspondence, starting with the works of
\cite{warn,bs}. Here, we will concentrate on the particularly
interesting cases of uniform distributions of D3-branes on a disc
and on a three-sphere.

\subsubsection{The disc}

The field-theory limit of the metric for $N$ D3-branes uniformly
distributed over a disc of radius $r_0$ reads
\ba
\label{4-8}
ds^2 &=& H^{-1/2} (- dt^2 + d \vec{x}_3^2 ) + H^{1/2}
{u^2+r_0^2 \cos^2\th \ov u^2+r_0^2}\ du^2
\nonumber\\
&+& H^{1/2}\left[(u^2+r_0^2\cos^2\th )d\th^2
+ r^2 \cos^2\th d\Om_3^2 + (u^2+r_0^2)\sin^2\th d \phi_1^2 \right]\ ,
\ea
where
\be
\label{4-9}
H = {R^4 \ov u^2 (u^2 + r_0^2 \cos^2 \th )}\ ,
\ee
while $d \Omega_3^2$ is the ${\rm S}^3$ metric.
Since the only
scale parameter entering into the supergravity solution is $r_0$
it is convenient to measure lengths and energies using this as a
reference scale. Setting $u \to r_0 u$ and $u_0 \to r_0 u_0$ and
introducing the dimensionless length and energy parameters by
\be
\label{4-11} L \to {R^2 \ov r_0} L\ ,\qq E \to {r_0 \ov \pi} E\ .
\ee
all dependence on $r_0$ and $R$ drops out so that we may set $r_0
\to 1$ and $R \to 1$ in what follows. The functions in \eqn{2-2}
and \eqn{2-3} now depend on $\th$ and read
\be
\label{4-12}
g(u,\th) = {u^2 + \cos^2\th \ov u^2 +1}\ ,\qq f_y(u,\th) = u^2(u^2+\cos^2\th)\ ,
\ee
and
\be
\label{4-13}
f_x(u,\th) = u^2(u^2+\cos^2\th)\ ,\qq f_\th(u,\th)
= u^2+\cos^2\th\ ,\qq h(u,\th) = {u^2 + \cos^2\th \ov u^2 +1}\ ,
\ee
respectively. The conditions \eqn{2-9} are satisfied only for
$\th_0=0$ and $\th_0=\pi/2$, which correspond to trajectories
orthogonal to the disc and lying on the plane of the disc,
respectively. To evaluate the quark-antiquark potential, we
examine these two trajectories in turn.

\no $\bullet$ $\th_0=0$. For this case, the integrals for the
dimensionless length and energy read \cite{bs}
\ba
\label{4-14}
L &=& 2 u_0 \sqrt{u_0^2 + 1} \int_{u_0}^\infty
 {du \ov u \sqrt{(u^2+1)(u^2-u_0^2)(u^2+u_0^2+1)}} \nonumber\\
&=& {2 u_0 k^{\prime} \ov u_0^2 +1} \left[ \elPi (k^{\prime 2},k) - \elK(k) \right]
\ea
and
\ba
\label{4-15}
E &=& \int_{u_0}^\infty du
\left[ u \sqrt{u^2+1 \ov (u^2-u_0^2)(u^2+u_0^2+1)} - 1 \right] - u_0 \nonumber\\
&=& \sqrt{2u_0^2+1} \left[ k^{\prime 2} \elK (k) - \elE(k) \right] \ ,
\ea
where $\elK(k)$, $\elE(k)$ and $\elPi(\a,k)$ denote the complete
elliptic integrals of the first, second and third kind
respectively and
\be
\label{4-16}
k={u_0 \ov \sqrt{2u_0^2+1}}\ , \qq k^{\prime}=\sqrt{1-k^2}\ ,
\ee
\begin{figure}[!t]
\begin{center}
\begin{tabular}{cc}
\includegraphics[height=5.2cm]{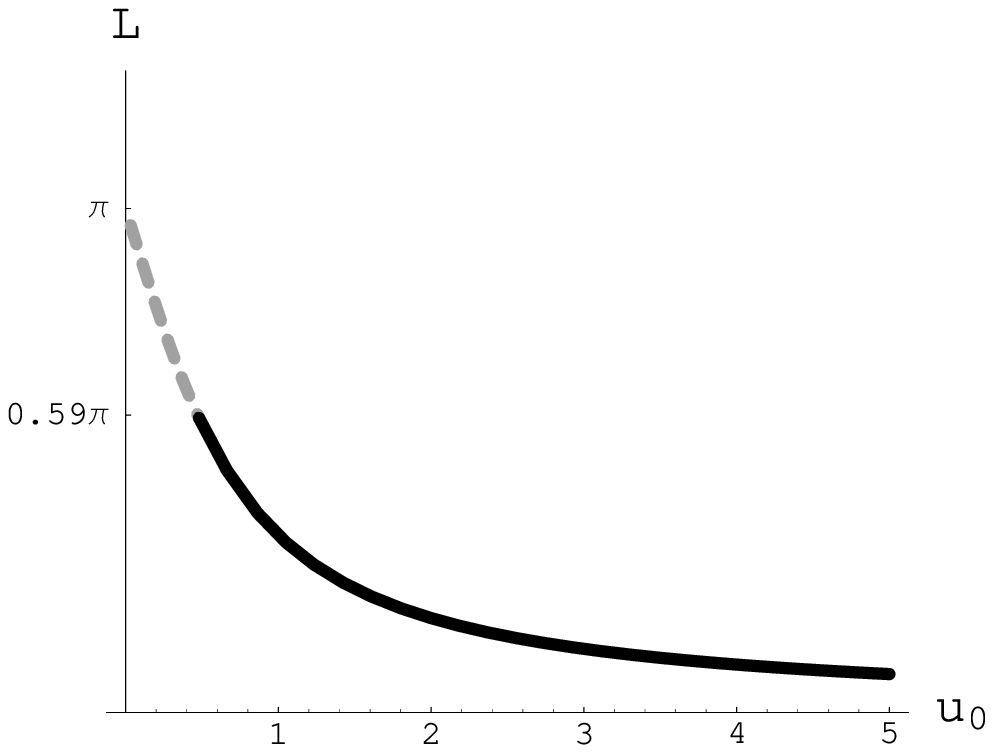}
&\includegraphics[height=5.2cm]{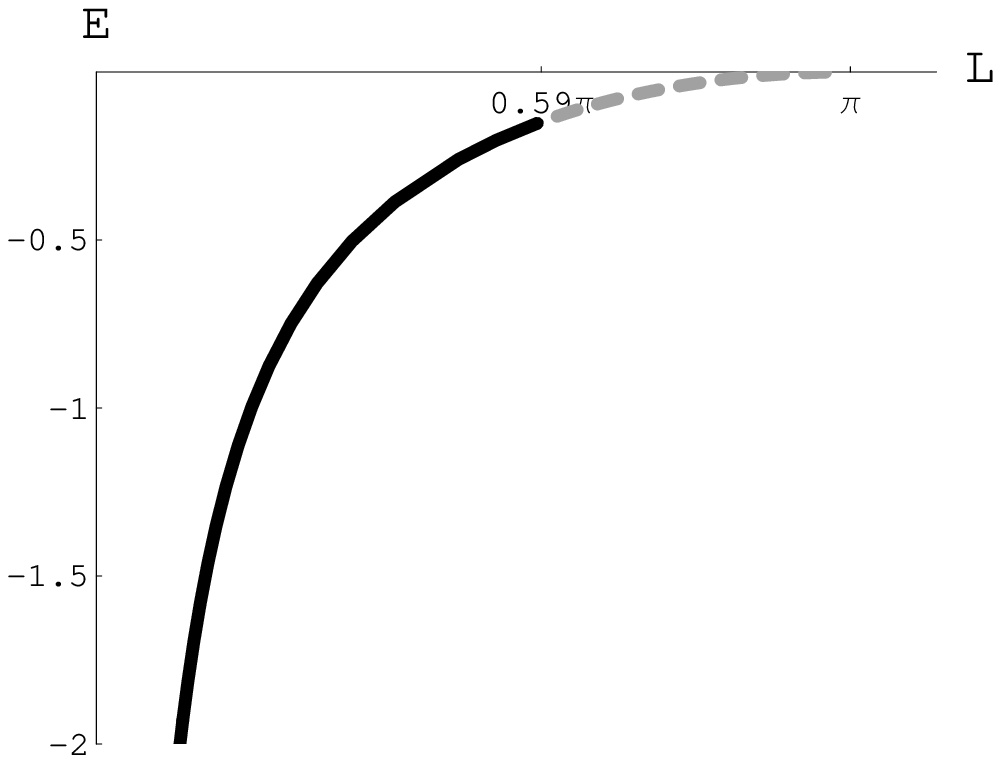}\\
(a) & (b)
\end{tabular}
\end{center}
\vskip -.5 cm \caption{Plots of $L(u_0)$ and $E(L)$ for the disc
at $\th_0=0$.} \label{fig2}
\end{figure}are the modulus and the complementary modulus. For $u_0 \gg 1$,
the behavior is Coulombic as before and at the opposite limit,
$u_0 \to 0$, we have the asymptotics \cite{bs}
\be
\label{4-17} L \simeq \pi(1-u_0)\ ,\qq E \simeq - {\pi \ov 4}
u_0^2\ ,
\ee
Now, $L(u_0)$ is a monotonously decreasing function and hence its
global maximum is at $u_{0 {\rm c}} = 0$ given by $L_{\rm c} =
\pi$. For $L > L_{\rm c}$, only the disconnected solution exists,
while for $L < L_{\rm c}$, Eq. \eqn{4-14} has a single solution
for $u_0$ and $E$ is a single-valued function of $L$. This
behavior is shown in the plots of Fig. \ref{fig2}, and it
corresponds to a screened Coulomb potential.

\no $\bullet$ $\th_0=\pi/2$. Now, the integrals for the
dimensionless length and energy read
\ba
\label{4-18}
L &=& 2 u_0^2 \int_{u_0}^\infty {du \ov u \sqrt{(u^2+1)(u^4-u_0^4)}} \nonumber\\
&=& {2 u_<^2 \ov \sqrt{u_0^2+u_>^2}}\left[ {\bf \Pi}
\left({u_>^2\ov u_0^2+u_>^2},k\right)-{\bf K}(k)\right ]
\ea
and
\ba
\label{4-19}
E &=& \int_{u_0}^\infty du u
\left[ {u^2 \ov \sqrt{(u^2+1)(u^4-u_0^4)}}
- {1 \ov \sqrt{u^2+1}} \right] - \int_0^{u_0} {du u \ov \sqrt{u^2+1}} \nonumber\\
&=&{u_0^2\ov \sqrt{u_0^2+u_>^2}} \ {\bf K}(k) - \sqrt{u_0^2+u_>^2} \ {\bf E}(k) + 1\ ,
\ea
where now
\be
\label{4-20}
k^2 = {u_>^2 -u_<^2 \ov u_0^2+u_>^2}\ , \qq k^{\prime}=\sqrt{1-k^2}\
\ee
\begin{figure}[!t]
\begin{center}
\begin{tabular}{cc}
\includegraphics[height=5.2cm]{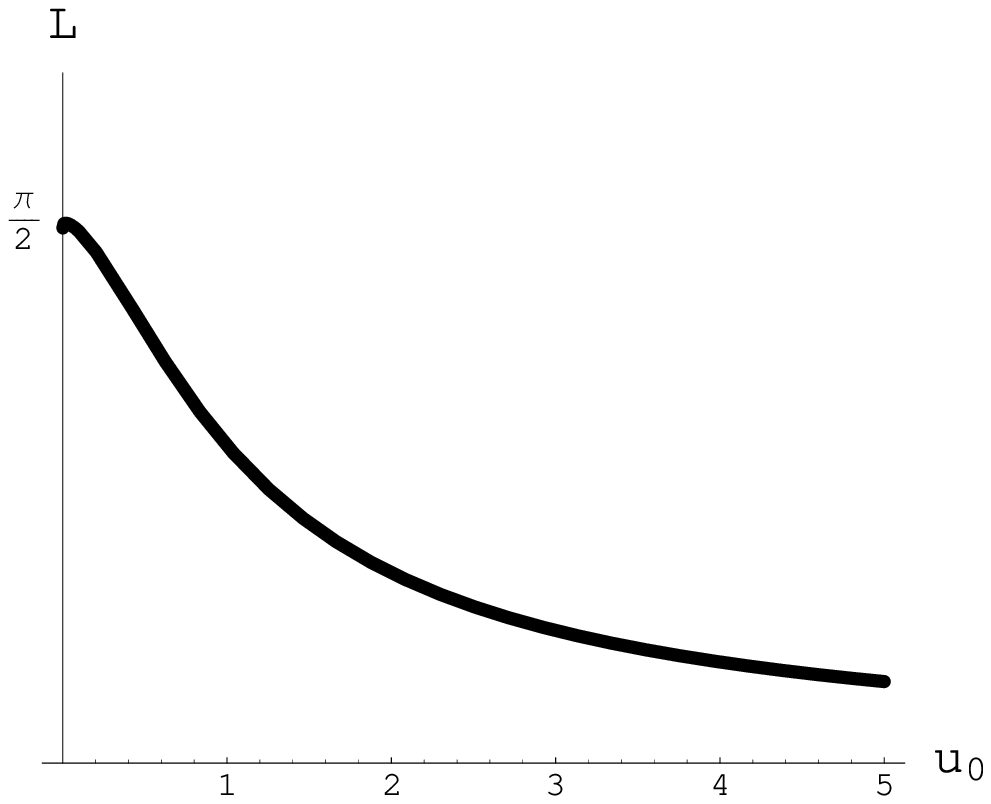}
&\includegraphics[height=5.2cm]{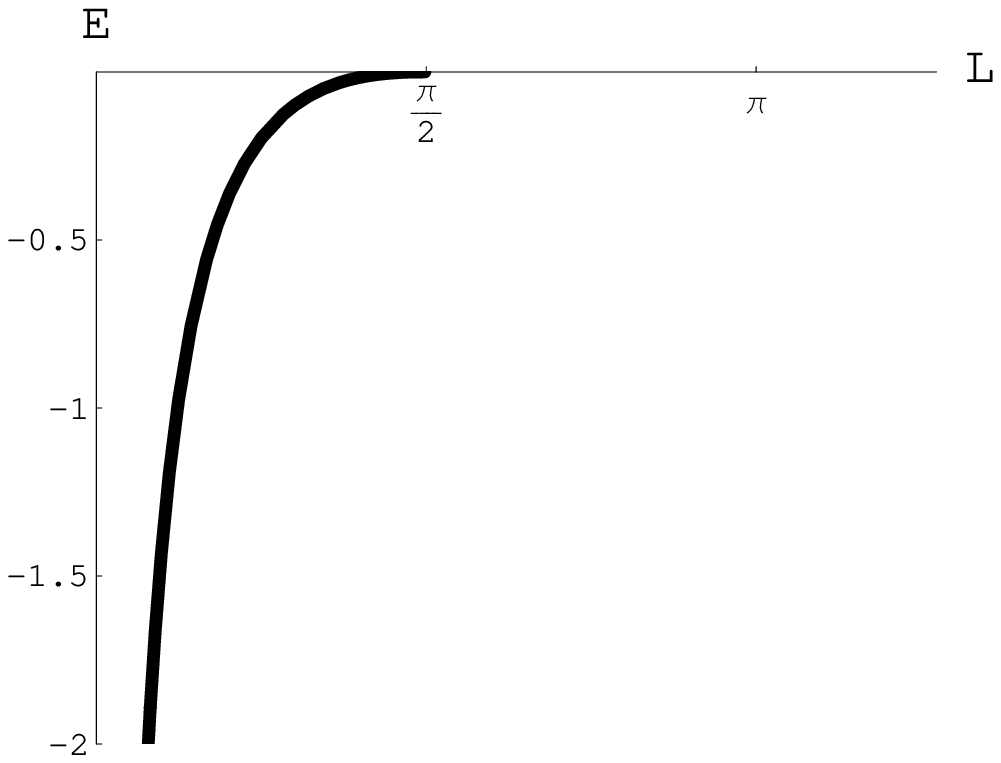}\\
(a) & (b)
\end{tabular}
\end{center}
\vskip -.5 cm \caption{Plots of $L(u_0)$ and $E(L)$ for the disc
at $\th_0=\pi/2$. Note that the screening length is less than the
the one for $\th_0=0$ by a factor of 2.} \label{fig3}
\end{figure}and $u_>$ ($u_<$) denotes the larger (smaller) between $u_0$ and
$1$. At the limit $u_0 \to 0$, we have the asymptotics \cite{bs}
\be
L \simeq {\pi\ov 2}\left[1- u_0^2 \left(\ln{8 \ov u_0^2}
-1\right)\right]\ , \qq E \simeq - {1\ov 8} u_0^4 \left(\ln {8 \ov
u_0^2}-{3\ov 2}\right)\ .
\ee
The behavior is qualitatively similar to the previous case with
the maximal length being
\be
\label{4-21} L_{\rm c} = {\pi \ov 2}\ .
\ee

\no Comparing the $\th_0=0$ and $\th_0=\pi/2$ cases we note that,
although the qualitative behavior of the potential is the same,
the expressions for the screening length differ by a factor of 2.
This factor is quite large as one expects that the orientation of
the string configuration will have a mild effect on physical
observables of the gauge theory. The apparent discrepancy will be
resolved by our stability analysis.

\subsubsection{The sphere}

The field-theory limit of the metric for $N$ D3-branes uniformly
distributed over a 3-sphere of radius $r_0$ reads
\ba
\label{4-22}
ds^2 &=& H^{-1/2} (- dt^2 + d \vec{x}_3^2 ) + H^{1/2}
{u^2-r_0^2 \cos^2\th \ov u^2-r_0^2}\ du^2
\nonumber\\
&+& H^{1/2}\left[(u^2-r_0^2\cos^2\th )d\th^2 + u^2 \cos^2\th d\Om_3^2
+ (u^2-r_0^2)\sin^2\th d \phi_1^2 \right]\ ,
\ea
where
\be
\label{4-23}
H = {R^4 \ov u^2 (u^2 - r_0^2 \cos^2 \th )}\ .
\ee
Note that this solution is obtained from the disc by taking $r_0^2
\to -r_0^2$. Employing the same rescalings as before, we write the
functions in \eqn{2-2} and \eqn{2-3} as
\be
\label{4-24}
g(u,\th) = {u^2 - \cos^2\th \ov u^2 - 1}\ ,\qq f_y(u,\th) = u^2(u^2-\cos^2\th)\
\ee
and
\be
\label{4-25}
f_x(u,\th) = u^2(u^2-\cos^2\th)\ ,\qq f_\th(u,\th)
= u^2-\cos^2\th\ ,\qq h(u,\th) = {u^2 - \cos^2\th \ov u^2 - 1}\ ,
\ee
respectively, and the conditions \eqn{2-9} are again satisfied
only for $\th_0=0$ and $\th_0=\pi/2$. We examine these two cases
in turn.

\no $\bullet$ $\th_0=0$. For this case, the integrals for the
dimensionless length and energy read \cite{bs}
\ba
\label{4-26}
L &=& 2 u_0 \sqrt{u_0^2 - 1} \int_{u_0}^\infty {du \ov u \sqrt{(u^2-1)(u^2-u_0^2)(u^2+u_0^2-1)}} \nonumber\\
&=& {2 u_0 k^{\prime} \ov u_0^2-1} \left[ \elPi (k^{\prime 2},k) - \elK(k) \right]
\ea
and
\ba
\label{4-27}
E &=& \int_{u_0}^\infty du
\left[ u \sqrt{u^2-1 \ov (u^2-u_0^2)(u^2+u_0^2-1)} - 1 \right] - (u_0-1) \nonumber\\
&=& \sqrt{2u_0^2-1} \left[ k^{\prime 2} \elK (k) - \elE(k) \right] + 1\ ,
\ea
where
\be
\label{4-28}
k={u_0 \ov \sqrt{2u_0^2-1}}\ ,\qq k^{\prime}=\sqrt{1-k^2}\ .
\ee
\begin{figure}[!t]
\begin{center}
\begin{tabular}{cc}
\includegraphics[height=5.2cm]{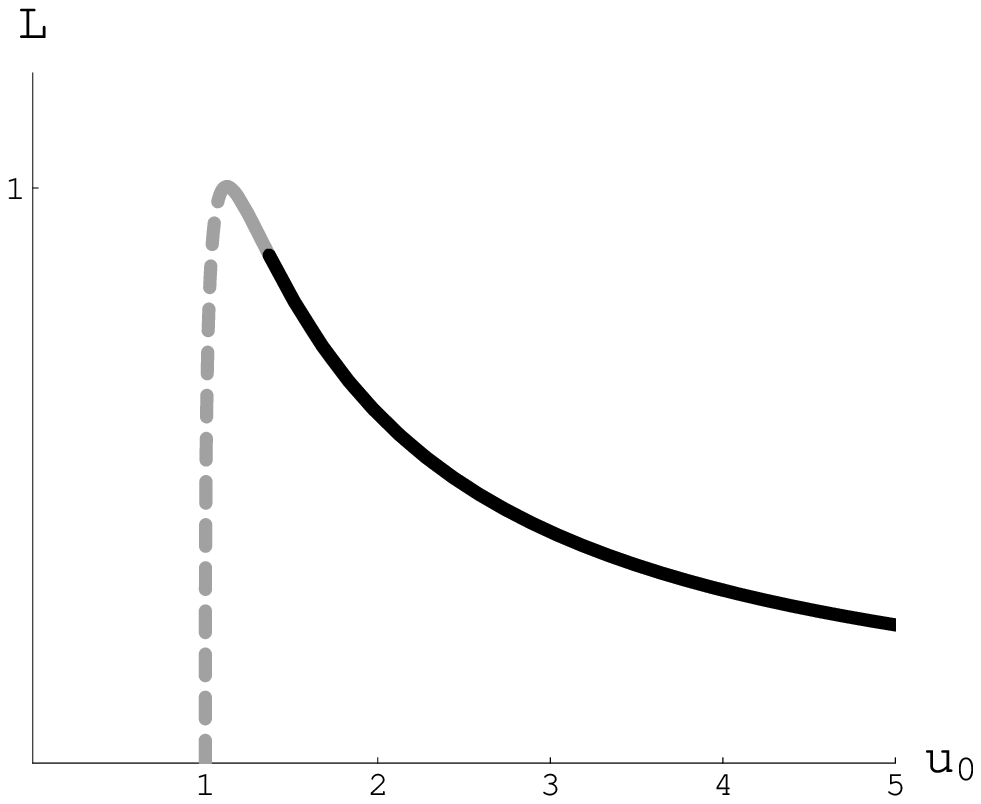}
&\includegraphics[height=5.2cm]{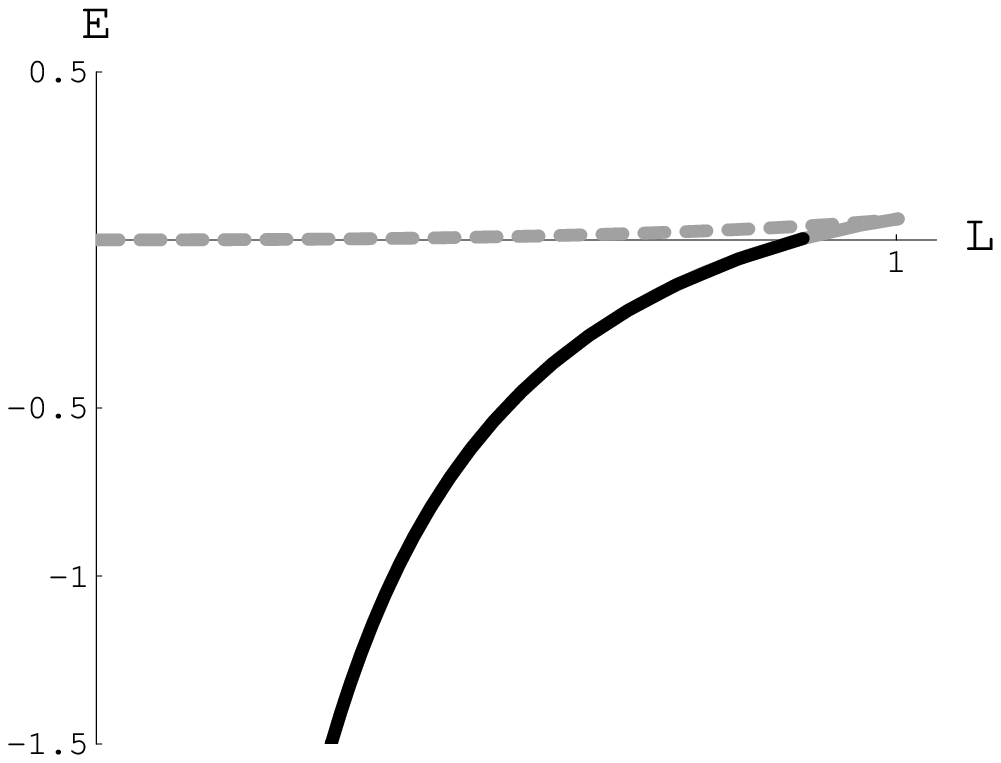}\\
(a) & (b)
\end{tabular}
\end{center}
\vskip -.5 cm \caption{Plots of $L(u_0)$ and $E(L)$ for the sphere
at $\th_0=0$. Note the appearance of two branches in the plot of
$E(L)$.} \label{fig4}
\end{figure}For $u_0 \gg 1$, the behavior is Coulombic, whereas in the
opposite limit, $u_0 \to 1$, we have the asymptotics \cite{bs}
\be
\label{4-29}
L \simeq \sqrt{2(u_0-1)}
 \left[ \ln \left({8 \ov u_0-1}\right) - 2 \right]\ ,
\qq E \simeq {u_0-1 \ov 2} \left[ \ln \left({8 \ov u_0-1}\right) - 3 \right]\ .
\ee
The function $L(u_0)$ has a single global maximum. Its location,
its value and the corresponding value of the energy are \cite{bs}
\be
u_{0 \rm c} \simeq 1.125\ , \qq  L_{\rm c} \simeq 1.002\ ,\qq E_c
\simeq 0.063\ . \label{4-8b}
\ee
For $L > L_{\rm c}$, only the disconnected solution exists. For $L
< L_{\rm c}$, Eq. \eqn{4-26} has two solutions for $u_0$ and $E$
is a double-valued function of $L$. This behavior is shown in the
plots of Fig. \ref{fig4} and, discarding the upper branch of
$E(L)$, it corresponds to a screened Coulomb potential.

\no $\bullet$ $\th_0=\pi/2$. Now, the integrals for the
dimensionless length and energy read
\be
\label{4-30}
L = 2 u_0^2 \int_{u_0}^\infty {du \ov u \sqrt{(u^2-1)
(u^4-u_0^4)}} = {\sqrt{2} \ov u_0} \left[ \elPi \left( \ha,k \right) -\elK(k) \right]\ ,
\ee
and
\be
\label{4-31}
E = \int_{u_0}^\infty {du u \ov \sqrt{u^2-1}}
\left( {u^2 \ov \sqrt{u^4-u_0^4}} - 1 \right) - \int_1^{u_0}
{du u \ov \sqrt{u^2-1}} = {u_0 \ov \sqrt{2}} \left[ \elK(k)- 2 \elE(k) \right]\ .
\ee
where now
\be
\label{4-32}
k=\sqrt{{u_0^2+1 \ov 2 u_0^2}}\ ,\qq k^{\prime}=\sqrt{1-k^2}\ .
\ee
\begin{figure}[!t]
\begin{center}
\begin{tabular}{cc}
\includegraphics[height=5.2cm]{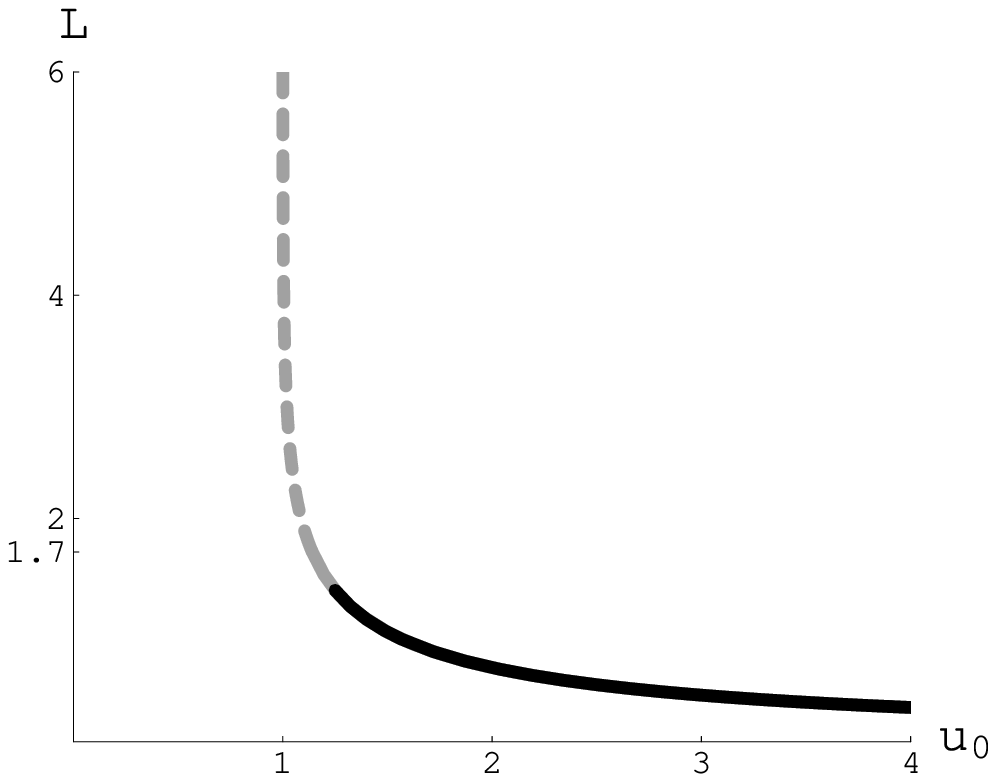}
&\includegraphics[height=5.2cm]{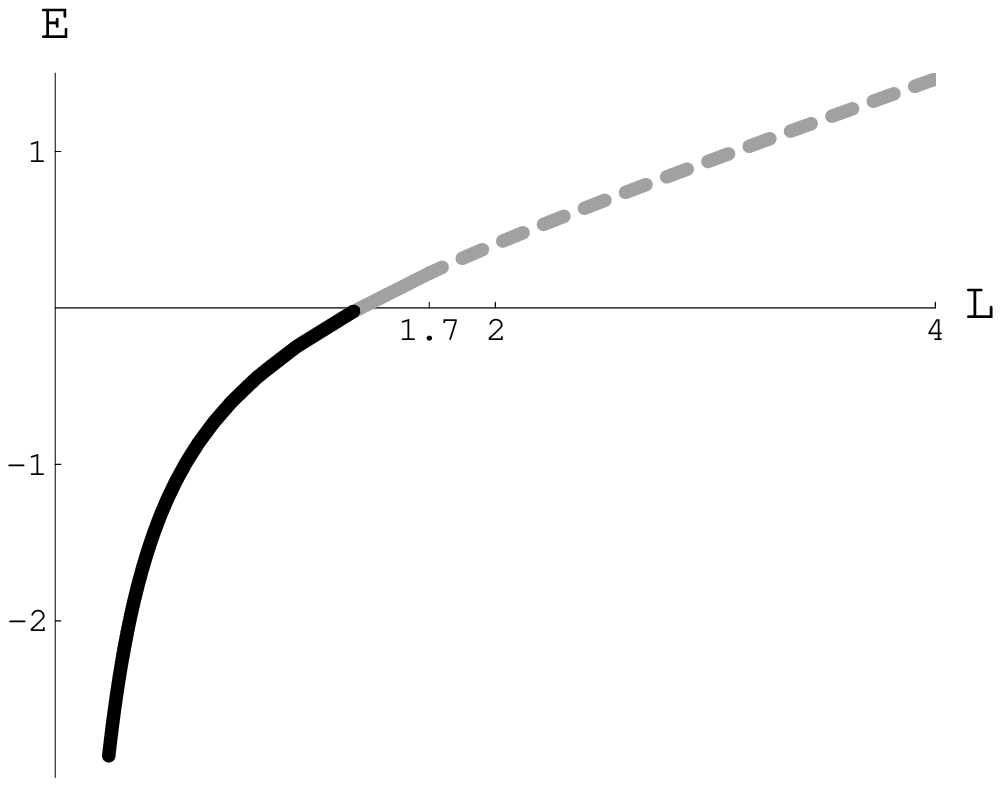}\\
(a) & (b)
\end{tabular}
\end{center}
\vskip -.5 cm \caption{Plots of $L(u_0)$ and $E(L)$ for the sphere
at $\th_0=\pi/2$.
 Note the appearance of a confining potential for large values of $L$.}
\label{fig5}
\end{figure}For $u_0 \gg 1$, the behavior is Coulombic, whereas in the
opposite limit, $u_0 \to 1$, we have the asymptotics \cite{bs}
\be
\label{4-33} L \simeq {1 \ov \sqrt{2}} \left[ \ln \left({16 \ov
u_0-1}\right) - 2\sqrt{2} \ln(1+\sqrt{2}) \right]\ , \qq E \simeq
{1 \ov 2 \sqrt{2}} \left[ \ln \left({16 \ov u_0-1}\right) - 4
\right]\ .
\ee
Now, $L(u_0)$ is a monotonously decreasing function which
approaches infinity as $u_0 \to 1$ and zero as $u_0 \to \infty$
and hence no maximal length exists, Eq. \eqn{4-30} has a single
solution for $u_0$, and $E$ is a single-valued function of $L$.
Therefore, it seems that no screening occurs but, instead, from
Eq. \eqn{4-33} we have a confining potential
\be
\label{4-34} E \simeq  {L \ov 2}\ ,\qq {\rm for} \quad L \gg 1 \ ,
\ee
whose appearance is quite puzzling, given that the underlying
theory is $\cN=4$ SYM.

\no Comparing the $\th_0=0$ and $\th_0=\pi/2$ cases, we note that
the qualitative behavior of the potential appears to be completely
different: in the first case the potential is a double-valued
function of $L$ which exists up to a maximal length, whereas in
the second case the potential exists for all values of the length
and interpolates between a Coulombic and a confining potential.
One does not expect such major qualitative differences to occur by
a simple change of the orientation of the string and, moreover,
the appearance of a confining potential is unexpected. Both of
these issues will be resolved by our stability analysis.

\section{D3-brane backgrounds: Stability analysis}

In this section we apply the stability analysis developed in
section 3 to the string solutions reviewed in section 4. We find
that for the non-extremal D3-brane and the sphere with $\th_0=0$
we have a longitudinal instability corresponding to the upper
branch of the energy curve, while for the disc with $\th_0=0$ and
for the sphere with $\th_0=\pi/2$ we have angular instabilities
towards the IR, even though the potential has a single branch.

\subsection{The conformal case}

As a simple example, and as a consistency check, let us first
consider the conformal case, corresponding to the $\mu \to 0$ or
$r_0 \to 0$ limit of any of the above solutions. In this case, the
Schr\"odinger potentials read
\ba
V_x(u;u_0) &=& 2 u^2 \ ,
\nonumber\\
V_\th(u;u_0) &=& 0\ ,\\
V_y(u;u_0) &=& 2 {u^{4} - u_0^4 \ov u^2}\ ,
\nonumber
\ea
where the subscript indicates the corresponding fluctuating
variable. The change of variables \eqn{3-8} explicitly gives
\be
x={1\ov u}\ {}_2F_1\left(\ha,{1\ov 4},{5\ov 4};{u_0^4\ov
u^4}\right)\ ,
\ee
and, hence, $x\in [0,x_0]$ with
\be
x_0 = {\G(1/4)^2\ov 4 \sqrt{2 \pi}}\ {1\ov u_0}\ .
\ee
Since all Schr\"odinger equations are defined in a finite interval
with positive-definite potentials, the corresponding energy
eigenvalues are positive and so there is no instability. The
equations corresponding to \eqn{3-7} for the transverse ($\d x$)
and longitudinal ($\d y$) fluctuations are given by Eq. (7) of
\cite{cg} and Eq. (16) of \cite{kmt} respectively. The comparison
for $\d x$ is immediate if we change variables as $u=1/z$ and also
rename $z_{\rm m}=1/u_0$, where $z$ and $z_{\rm m}$ are the
variable and parameter used in \cite{cg}. For $\d y$, the
comparison involves the same change of variables as before and $\d
y = u_0^2 u^2 (u^4-u_0^4)^{-1/2} \d
\bar y$, where $\d \bar y$ is the variable called $u$ in
\cite{kmt}. In addition, since the independent variable in Eq.
(16) of \cite{kmt} is $y$ ($x$ in their notation) we should also
use the differential relation $dy = u_0^2/u^2 (u^4-u_0^4)^{-1/2}
du$, resulting from the classical equation of motion \eqn{2-11}.

\subsection{Non-extremal D3-branes}

We next proceed to the case of non-extremal D3-branes, where we
recall that the potential energy is a double-valued function of
the separation length. The Schr\"odinger potentials for the three
types of fluctuations are given by
\ba
\label{abc}
V_x(u;u_0) &=& 2 {u^8 - u_0^4 \ov u^6}\ ,\nonumber\\
V_\th(u;u_0) &=& 0\ ,\\
V_y(u;u_0) &=& 2 {u^{12} - u_0^4 u^8 - (4u_0^4 -3 ) u^4 + u_0^4
\ov u^6 (u^4-1)}\ .
 \nonumber
\ea
The value of the endpoint $x_0$ is found using \eqn{hj1}
and reads
\be
x_0 = {\G(1/4)^2\ov 4 \sqrt{2 \pi}}\ {1\ov u_0}\
{}_2F_1\left(\ha,{1\ov 4},{3\ov 4};{1\ov u_0^4}\right)\ ,
\ee
with the behaviors \eqn{uvx0} and
\ba
x_0(u_0)\simeq -{1\ov 4} \ln(u_0-1) +{\pi +\ln 64\ov 8} + {\cal
O}(u_0-1)\ .
\label{5-6}
\ea
Since $V_x$ is positive for all values of $u_0$, the solution is
stable under transverse perturbations, in accordance with our
general result in section 3.3. Also, $V_\th$ is identically zero,
which means that, given that $x_0$ is finite, the spectrum is
positive-definite and the solution is stable against angular
perturbations as well. On the other hand, $V_y$ starts from a
negative value at $u=u_0$ given by
\ba
V_{y 0} = -{8\ov u_0^2} \ ,\qq -8 \leqslant V_{y 0} < 0\
\ea
and behaves as in \eqn{3-31} at $u\to \infty$. To examine the
occurrence of instabilities under longitudinal perturbations, we
apply the results of section 3.3 and the perturbation-theory
formulas of section 3.4 to our problem. We find that there is a
longitudinal zero mode at $u_0=u_{0 \rm c}$ where $u_{0 \rm c}$ is
found by numerically solving the equation \eqn{3-30} which, for
our case, reads
\ba
{9 \ov 5 u_0^4} \ {}_2F_1\left(\ha ,{7\ov 4},{9\ov 4};{1\ov
u_0^4}\right) = {}_2F_1\left(\ha ,{3\ov 4},{5\ov 4};{1\ov
u_0^4}\right) \ .
\ea
This has precisely one solution given by the first of \eqn{4-8a},
i.e. $u_{0\rm c}\simeq 1.177$. To find the change in the energy
eigenvalue $\omega^2$ as we move away from $u_{0 \rm c}$, we first
use \eqn{3-28} to find the explicit expression for the
longitudinal zero mode
\ba
\d y = {1.463\ov u^3}\ F{}_1\left({3\ov 4}; {3\ov 2} , -{1\ov 2} ,
{7\ov 4} ;{u_{0\rm c}^4\ov u^4}, {1\ov u_{0\rm c}^4}\right)\ ,
\ea
where $F{}_1(a\ ;b_1,b_2,c\ ;x,y)$ is the Appell hypergeometric
function and the normalization has been fixed from the
corresponding Schr\"odinger wavefunction which reads
\be
\Psi = {1.463 \ov u^2} \sqrt{u^4-u_{0\rm c}^4\ov u^4-1} \
F{}_1\left({3\ov 4}; {3\ov 2} , -{1\ov 2} , {7\ov 4} ;{u_{0\rm
c}^4\ov u^4}, {1\ov u_{0\rm c}^4}\right)\ .
\ee
Then, using Eq. \eqn{3-22}, we find that the change in energy
eigenvalue is given by
\be
\d \omega^2 \simeq ( 51.9 + 29.6 - 20.1 )\ \d u_0 = 61.4\ \d u_0 \
,
\ee
where each term inside the parentheses represents the contribution of
the corresponding term in \eqn{3-22}. Hence, as we move to the
right (left) of $u_{0{\rm c}}$, the eigenvalue $\omega^2$ becomes
positive (negative) and therefore the long string is unstable
under longitudinal perturbations. This is what is expected from
the energetics of these configurations, shown in Fig. 1(b).

\subsection{Multicenter D3-branes}

We finally turn to the case of multicenter D3-branes, where all
types of problematic behavior discussed in section 2 appear. In
what follows, we present the results of the stability analysis for
all cases discussed in section 4.2.

\subsubsection{The disc}

For the disc distribution, we recall that both allowed
orientations of the string lead to a screened Coulomb potential,
but the screening lengths differ by a factor of 2. The results of
the stability analysis for the two orientations are as follows.

\no
$\bullet$ $\th_0=0$. In this case, we have the Schr\"odinger
potentials
\ba
V_x(u;u_0) &=& {8 u^8 + 18 u^6 + 11 u^4 - [2 u_0^2(u_0^2+1)-1]
 u^2 + u_0^2(u_0^2+1) \ov 4 u^2(u^2+1)^2}\ ,\nonumber\\
V_\th(u;u_0) &=& - {2 u^6 + u^4 + [6 u_0^2(u_0^2+1)-1] u^2 + 3
 u_0^2(u_0^2+1) \ov 4 u^2(u^2+1)^2}\ ,\\
V_y(u;u_0) &=& {8 u^8 + 18 u^6 - [ 8 u_0^2 (u_0^2+1) - 11] u^4
- [6 u_0^2(u_0^2+1)-1] u^2 -3 u_0^2(u_0^2+1) \ov 4 u^2(u^2+1)^2}\ ,
\nonumber
\ea
and the value of the endpoint $x_0$ reads
\ba
x_0 = \sqrt{2 {k'}^2 -1} \ {\bf K}(k')\ ,
\ea
where $k'$ is the complementary modulus defined in \eqn{4-16}. We
have the general behavior \eqn{uvx0} and
\ba
x_0(u_0)\simeq -\ln {u_0\ov 4} + {\cal O}(u_0^2\ln u_0)\ .
\label{xo1}
\ea
Since $V_x$ and $V_y$ are positive for all values of $u_0$, the
solution is stable against transverse and longitudinal
perturbations, in accordance with the general results of section
3.3 in the absence of a maximal length. On the other hand, $V_\th$
is negative throughout the whole range of $u_0$, with its values
at $u=u_0$ and $u\to \infty$ given by
\ba
V_{\th 0} &=& -2 +{3\ov 2(u_0^2+1)} \ ,\qq -2 < V_{\th 0}
\leqslant -\ha \ ,
\nonumber\\
V_{\th\infty} &=& -\ha\ .
\label{5-6n}
\ea
To examine the occurrence of instabilities, we use the
infinite-well approximation of \eqn{kwsft1} and we examine the
behavior of the lowest eigenvalue $\omega_0^2$ by plotting it as a
function of $u_0$. We find that it is an increasing function
starting at negative values for $u_0=0$ and changing sign at a
critical value $u_{0\rm c}$. This value, and the corresponding
maximal length and energy, are given by
\ba
u_{0\rm c}\simeq 0.48\ ,\qq L_{\rm c} \simeq 0.59 \pi\ ,\qq E_{\rm
c} \simeq  -0.15 \ . \label{5-7}
\ea
Therefore, when the separation distance of the quark-antiquark
pair becomes larger than the value $L_{\rm c}$ given above, small
fluctuations in $\th$ destabilize the corresponding classical
solution and the resulting potential should not be trusted. The
true screening length is thus given by the second of \eqn{5-7} and
turns out to be comparable to that for $\th_0=\pi/2$. Moreover,
within the infinite-well approximation, we may show that for
$u_0<u_{0\rm c}$ more states become negative. Indeed, setting
$\omega^2_n(u_0)=0$ and using the limiting behavior \eqn{xo1}, we
find that the value $u_{0 {\rm c},n}$ in which the $n$--th energy
eigenvalue becomes zero is given by the formula
\ba
u_{0 {\rm c},n} \simeq 4 e^{-\pi (2n+1)/\sqrt{2}}\ ,
\ea
valid in practice for all $n\geqslant 1$.

\no $\bullet$ $\th_0=\pi/2$. Now, the Schr\"odinger potentials are
given by
\ba
V_x(u;u_0) &=& {2 u^6 + u^4 + u_0^4 \ov u^4}\ ,
\nonumber\\
V_\th(u;u_0) &=& 1\ ,\\
V_y(u;u_0) &=& {2 u^6 + u^4 - 2 u_0^4 u^2 - 3u_0^4 \ov u^4}\ ,
\nonumber
\ea
and the value of the endpoint $x_0$ reads
\ba
x_0 = {1\ov \sqrt{u_0^2 +u_>^2}}\ {\bf K}(k)\ ,
\ea
where $k$ is the modulus defined in \eqn{4-20}.
Its behavior is given by \eqn{uvx0} and
\be
x_0(u_0)\simeq - \ln {u_0 \ov \sqrt{8}} + {\cal O}(u_0^4 \ln u_0)\
.
\ee
Since $V_x$ and
$V_\th$ are manifestly positive, the solution is stable under
transverse and angular perturbations. Also, although $V_y$ has a
negative part below some value for $u$, the fact that no critical
points of the length exist in this case implies that the solution
is stable under longitudinal perturbations as well.

\no The upshot of this analysis is that the screening lengths for
the two different extreme angles for which the heavy quark
potential can be computed become practically the same which is a
requirement for the notion of a screening length to make physical
sense. It is natural to expect that, if the system starts with
$\th_0=0$ and $L_c < L < \pi$, small fluctuations will tend to
drive the value of $\th$ towards $\pi/2$.

\subsubsection{The sphere}

For the sphere distribution, we recall that the two allowed
orientations of the string lead to quite different behaviors and
that a confining potential appears in the $\th_0=\pi/2$ case. The
results of the stability analysis for the two orientations are as
follows.

\no
$\bullet$ $\th_0=0$. In this case, we have the Schr\"odinger potentials
\ba
V_x(u;u_0) &=& {8 u^8 - 18 u^6 + 11 u^4 + [2 u_0^2(u_0^2-1)-1]
 u^2 + u_0^2(u_0^2-1) \ov 4 u^2(u^2-1)^2}\ ,\nonumber\\
V_\th(u;u_0) &=& {2 u^6 - u^4 + [6 u_0^2(u_0^2-1)-1] u^2 - 3
u_0^2(u_0^2-1) \ov 4 u^2(u^2-1)^2}\ ,
\label{5-14}\\
V_y(u;u_0) &=& {8 u^8 - 18 u^6 - [ 8 u_0^2 (u_0^2-1) - 11] u^4
 + [6 u_0^2(u_0^2-1)-1] u^2 -3 u_0^2(u_0^2-1) \ov 4 u^2(u^2-1)^2}\ ,\nonumber
\ea
and the value of the endpoint $x_0$ reads
\ba
x_0 = \sqrt{1-2 {k'}^2}\ {\bf K}(k')\ ,
\ea
where $k'$ is the complementary modulus defined in \eqn{4-28}. We
have the general behavior \eqn{uvx0} and
\ba
x_0(u_0)\simeq {\pi\ov 2} -{3\pi\ov 4} (u_0-1) + {\cal O}(u_0-1)^2\ .
\label{xo2}
\ea
Since $V_x$ and $V_\th$ are positive for all values of the
parameter $u_0$, the solution is stable under transverse and
angular perturbations. On the other hand, $V_y$ starts from a
negative value and turns positive. Repeating the analysis of
section 5.1, we find that a longitudinal instability occurs for
$u_0$ below the the critical value $u_{0 \rm c}$ given in the
first of \eqn{4-8b}.

\no $\bullet$ $\th_0=\pi/2$. Now, the Schr\"odinger potentials are
given by
\ba
V_x(u;u_0) &=& {2 u^6 - u^4 - u_0^4 \ov u^4}\ ,\nonumber\\
V_\th(u;u_0) &=& -1\ ,\\
V_y(u;u_0) &=& {2 u^6 - u^4 - 2 u_0^4 u^2 + 3u_0^4 \ov u^4}\ .
\nonumber
\ea
and the value of the endpoint $x_0$ reads
\ba
x_0 = \sqrt{2 k^2-1\ov 2}\ {\bf K}(k)\ ,
\ea
where $k$ is the modulus defined in \eqn{4-32}.  We have the
general behavior \eqn{uvx0} and
\ba
x_0(u_0)\simeq -{1\ov 2\sqrt{2}}\ln {u_0-1\ov 16}  + {\cal
O}\left((u_0-1)\ln(u_0-1)\right)\ . \label{xo3}
\ea
Since $V_x$ and $V_y$ are positive for all values of $u_0$, the
solution is stable against transverse and longitudinal
perturbations, again in accordance with the general results of
section 3.3. On the other hand, $V_\th$ has a constant negative
value which in particular implies that the infinite-well
approximation is exact. Examining the behavior of the lowest
eigenvalue $\omega_0^2$, we find that it is an increasing function
starting at negative values for $u_0=0$ and changing sign at a
critical value $u_{0\rm c}$ with
\ba
u_{0\rm c}\simeq 1.14\ ,\qq L_{\rm c} \simeq 1.7\ ,\qq E_{\rm c} \simeq
0.22 \ .
\label{5-7c}
\ea
That is, when the quark-antiquark separation becomes larger than
$L_{\rm c}$, small fluctuations in $\th$ destabilize the classical
solutions and the resulting potential should not be trusted. Since
the configurations giving rise to a linear potential correspond to
separations larger than $L_{\rm c}$, the confining behavior is
spurious and instead we have a screened Coulomb potential with
screening length given by the second of \eqn{5-7c}. Moreover, we
may show that for $u_0<u_{0\rm c}$ more states become negative.
Indeed, setting $\omega^2_n(u_0)=0$ and using \eqn{xo3} we find
that the value $u_{0 {\rm c},n}$ in which the $n$--th energy
eigenvalue becomes zero is
\ba
u_{0 {\rm c},n} \simeq 1+ 16\ e^{-\sqrt{2}\pi (2n+1)}\ ,
\ea
valid, practically, for all $n\geqslant 1$.

\no The upshot of this analysis is that there is no stable
confining branch and that both potentials are of the screened
Coulomb type, with comparable screening lengths.

\subsection{Special points}

We have seen that, once we cross the critical value for $u_0$ from
above, the lowest eigenvalue of the fluctuations in the
$y$-direction becomes negative and the only way for it to turn
back to positive values is the appearance of another extremum of
the length at a different value of $u_0$. However, as we have
already mentioned the singularity structure of the fluctuation
equations \eqn{3-6} changes when $u_0=u_{\rm min}$. This isolated
point corresponds to zero length and energy and to two straight
strings stuck together. It is easily seen that we can have a
positive-definite spectrum of fluctuations and therefore
perturbative stability. However, for fluctuations with a parameter
$u_0$ infinitesimally larger than $u_{\rm min}$, the spectrum has
a single negative eigenvalue. As we shall see, the apparent
paradox is resolved by the fact that perturbation theory breaks
down when applied to points in the vicinity of $u_0=u_{\rm min}$.
This shows that the special points with $u_0=u_{\rm min}$ are
really of measure zero in all physical processes, in the sense
that no conclusion reached at these points remains approximately
correct when we move, even infinitesimally, away from them.

\no This can be easily argued for the case of the longitudinal
fluctuations for the non-extremal D3-brane. When $u_0=1$ we see
from \eqn{abc} that the potential becomes
\be
V_y(u;1)=2 {u^8-1\ov u^6}\ ,\qq u\geqslant 1\ ,
\ee
whereas from \eqn{5-6} we see that the Schr\"odinger equation is
defined in the entire positive half-line. Hence the spectrum of
fluctuations is positive and any perturbative analysis around
$u_0=1$ necessarily breaks down.

\no A similar argument holds for the case of angular fluctuations
for the disc and the trajectory corresponding to $\th_0=0$. In
that case the Schr\"odinger problem can be expressed explicitly in
terms of the new variable $x$ in \eqn{3-8} as
\ba
u={1\ov\sinh x}\ , \qq  0\leqslant x < \infty\ .
\ea
The Schr\"odinger potential in terms of the variable $x$ is given
by
\ba
V_\th(x;0)={1\ov 4}\left(1-{3\ov \cosh^2 x }\right)\ .
\ea
It can be shown that the solution is given in terms of
hypergeometric functions and that the spectrum is continuous with
a gap, i.e. $\omega^2 > 1/4$. Since we know from the analysis
above that the spectrum is actually negative below the value given
approximately in \eqn{5-7}, we conclude, as before, that the
perturbative analysis around $u_0=0$ necessarily breaks down.

\no Next, we demonstrate explicitly all details of this phenomenon
in the particular case of the Coulomb branch for the sphere and
for the trajectory with $\th_0=0$, in which case longitudinal
fluctuations are unstable. If $u_0=1$, the Schr\"odinger potential
of the longitudinal fluctuations can be expressed explicitly as a
function of the variable $x$ of \eqn{3-8} which, for our case,
reads
\ba
u={1\ov\sin x}\ , \qq  0\leqslant x\leqslant {\pi\ov 2}\ .
\ea
it is given by
\ba
V_y(x;1)={2\ov \sin^2 x}-{1\ov 4 \cos^2 x}-{1\ov 4} \ ,
\ea
and falls into the class of P\"oschl--Teller potentials of type I.
The corresponding differential equation has a complete set of
orthogonal solutions given by
\ba
\Psi_n(x) = \sqrt{4n+5} \sin^2\!x \cos^{1/2}\!x \ P_n^{(3/2, 0)}(\cos 2x)\ ,\qq n=0,1,\dots \ ,
\label{ff}
\ea
where $P_n^{(\a,\b)}$ are the Jacobi polynomials of $n$--th order.
The respective eigenvalues are
\ba
\omega^2_n=4n^2+10n+6\ ,\qq n=0,1,2,\dots\ . \label{5-25}
\ea
This is a positive-definite spectrum, showing that small
fluctuations do not destabilize this special point at which
$u_0=1$, $L=0$ and $E=0$. Consider next a small deviation from the
value $u_0=1$. We will show that using \eqn{3-21} to compute the
correction to the potential leads to divergent integrals for the
corrections to the energy eigenvalues in \eqn{5-25}. For this, we
need the expressions
\ba
{\partial V_y\ov\partial u_0}\bigg |_{u_0=1} & = & \left(3-{4\ov
\sin^2 x}-{3\ov 2} \sin^2 x\right)\tan^4 x\ ,
\nonumber \\
{\partial V_y \ov \partial u}\bigg |_{u_0=1} & = & {1\ov 2 \sin
x}(8+\tan^4 x)\ ,
\\
{\del x_0 \ov \del u_0}\bigg |_{u_0=1}  & = & -{3\ov 4 \pi}\ ,
\nonumber
\ea
where we have used \eqn{5-14} and \eqn{xo2} (equivalent to
computing the integral in \eqn{j1} with $u_0=1$). From these we
may explicitly compute the right hand side of \eqn{3-21}, which is
of the form
\ba
\d V = (u_0-1) v(x)\ ,\qq v(x)\simeq {1\ov (x-\pi/2)^6} + {\cal O}\left( (x-\pi/2)^{-4}\right)\ ,
\ea
where $v(x)$ is a complicated function, with the indicated
singular behavior. From the expression \eqn{ff} for the solutions
to the unperturbed problem we find that, near $x=\pi/2$, $\Psi
\sim (\pi/2-x)^{1/2}$. Hence the integral $\int_0^{\pi/2} dx\
\Psi^2 \d V$ diverges, thus proving the breakdown of perturbation
theory near $u_0=1$.

\section{Discussion}

In this paper we have examined the perturbative stability of
string configurations dual to flux tubes between static
quark-antiquark pairs in $\cN=4$ SYM at finite temperature and at
the Coulomb branch. The motivation for our study was the fact that
the quark-antiquark potentials computed via the AdS/CFT
prescription for the above cases exhibit behaviors that are
inconsistent with our field-theory expectations, namely (i)
multiple branches of the potential, (ii) a heavily
orientation-dependent screening length, and (iii) a linear
confining behavior. Our stability analysis resolves the
discrepancy by showing that the configurations corresponding to
the upper branches of the potential are unstable against
longitudinal perturbations while those giving rise to an
orientation-dependent screening length and to a confining behavior
are unstable under angular perturbations.

\no The methods developed here can be extended to the more
involved situation of string configurations in a boosted and/or
rotating non-extremal D3-brane background, with the boost
corresponding to a thermal medium moving with respect to the pair
and the rotation corresponding to R-charge chemical potentials in
the gauge theory. In such cases, the metrics are no longer
diagonal and the various fluctuations are not guaranteed to
decouple so that the general discussions of sections 2 and 3 must
be modified. Nevertheless, we believe that an analytic treatment
in these cases is also possible.

\no Another potential application of these methods refers to
Wilson-loop calculations in less supersymmetric backgrounds. In
fact, there are several examples \cite{hsz,ahn-poritz} where
calculations of the heavy quark-antiquark potential in backgrounds
with $\cN=1$ supersymmetry yield, at large separations, a linear
confining behavior which, in contrast to the $\cN = 4$ case, is
actually expected on physical grounds. Therefore, it would be
particularly interesting to investigate whether the string
configurations giving rise to such a behavior are stable.

\vskip 0.5cm

\centerline{ \bf Acknowledgments}

\no We thank K.~Anagnostopoulos for helpful discussions.
K.~Sfetsos thanks U.~Wiedemann for related discussions during the
early stages of this project in July-August of 2006. K.~Sfetsos
and K.~Siampos acknowledge support provided through the European
Community's program ``Constituents, Fundamental Forces and
Symmetries of the Universe'' with contract MRTN-CT-2004-005104,
the INTAS contract 03-51-6346 ``Strings, branes and higher-spin
gauge fields'', the Greek Ministry of Education programs $\rm \P
Y\Th A\G OPA\S$ with contract 89194 and the program $\rm E\Pi A N$
with code-number B.545. K.~Siampos also acknowledges support
provided by the Greek State Scholarship Foundation (IKY).

\appendix

\section{An analog from classical mechanics}

It is not often appreciated that the problem of calculating Wilson
loops in the supergravity approach, especially in cases where
multiple branches of the solution appear, has striking
similarities to a textbook problem in classical mechanics, namely
that of determining the shape of a thin soap film stretched
between two rings (Plateau's problem).\footnote{See, however,
\cite{gross} for a discussion of this analogy in the context of
Wilson-loop correlators.} The main similarity of the two problems
lies in the fact that, although the solution of the equations of
motion is straightforward, the boundary conditions allow for
multiple solutions and introduce a phase structure. Since a lot of
insight for our problem can be gained by looking at this simpler
situation, in what follows we give a modern pedagogical review of
this mechanical analog. Details can be found in standard textbooks
on variational methods (e.g. \cite{variations}), while a partial
stability analysis has been done in \cite{durand}.

\no We consider a thin soap film stretched between two coaxial
circular rings of unit radius, separated by a distance $L$.
Neglecting gravity, we write the action as
\ba
\label{a-1}
S = \int dt \int d\s_1 d\s_2 \sqrt{\g}
\left[ {1 \ov 2}  ( \dot{x}_1^2 + \dot{x}_2^2 + \dot{x}_3^2) - 1 \right] \ ,
\ea
where we have taken the mass density and the surface tension equal
to $1$ and $\ha$ respectively. Here, $(x_1,x_2,x_3)$ are Cartesian
coordinates on the embedding space, while $(\s_1,\s_2)$ and
$\g_{\alpha\beta}$ are the coordinates and the induced metric on
the surface respectively. For a static, axially-symmetric
configuration we introduce cylindrical coordinates $(r,\phi,z)$ in
the embedding space, we use reparametrization invariance to set $(\s_1,\s_2) = (z,\phi)$
and we choose
the embedding $r=r(z)$. Then the action reduces to
\ba
\label{a-2} S = - 2\pi \int dt \int_{-L/2}^{L/2} dz r \sqrt{1 +
r^{\prime 2}}\ .
\ea
Independence of the Lagrangian from $z$ leads to the first
integral
\ba
\label{a-3} {r \ov \sqrt{1 + r^{\prime 2}}} = u_0\ ,
\ea
where $u_0$ is the value of $r$ at the point where $r^\prime(z)=0$
which by symmetry occurs at $z=0$. Integrating \eqn{a-3} and
imposing $r^\prime(0)=0$, we obtain the solution $r(z) = u_0 \cosh
(z / u_0)$, first found by Euler and defining a surface of
revolution known as the catenoid. The integration constant $u_0$
is specified by the boundary condition $r(\pm {L \ov 2})=1$, which
gives
\be
\label{a-5} L = 2 u_0 \cosh^{-1} {1 \ov u_0}\ .
\ee
The potential energy of the solution is
\ba
\label{a-6} E =\pi u_0 L \left( 1+ {u_0 \ov L} \sinh {L \ov u_0}
\right) = 2\pi \left( \sqrt{1-u_0^2} + u_0^2 \cosh^{-1} {1 \ov
u_0} \right)\ .
\ea
Note that these equations require $0 \leqslant u_0 \leqslant 1$ .
Besides the catenoid solution just discussed, there also exists
the so-called Goldschmidt solution which describes two
disconnected circular films on the two rings,\footnote{The
existence of this solution is most clearly seen by using the
parametrization $(\s_1,\s_2) = (r,\phi)$ and $z=z(r)$, for which
the first-order equation reads $ r z^{\prime} / \sqrt{1 +
z^{\prime 2}} = \const$ which is solved by $z^{\prime}=0$.} with
energy $E=2\pi$.
\begin{figure}[!t]
\begin{center}
\begin{tabular}{ccc}
\includegraphics[height=5cm]{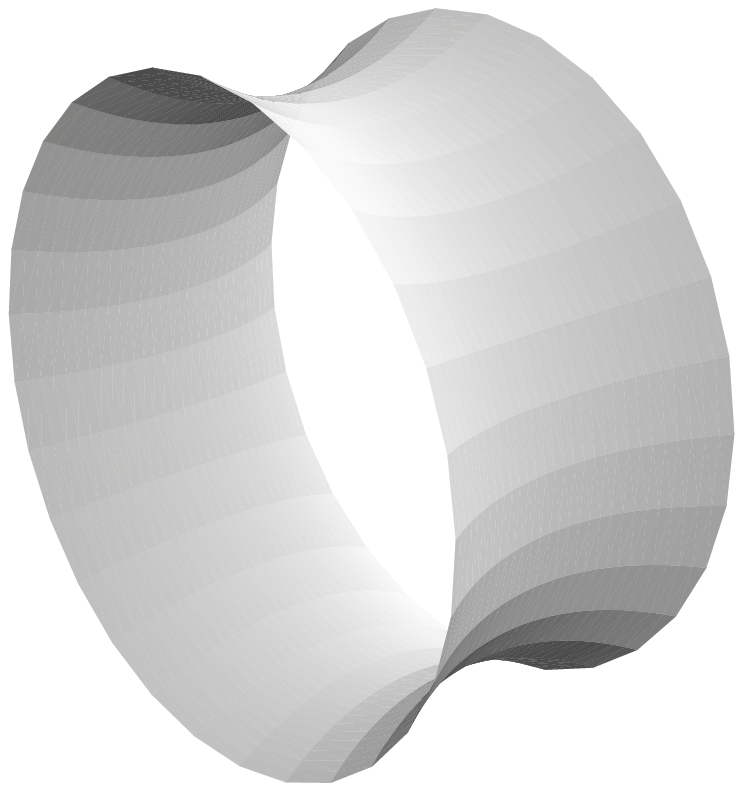}
&\includegraphics[height=5cm]{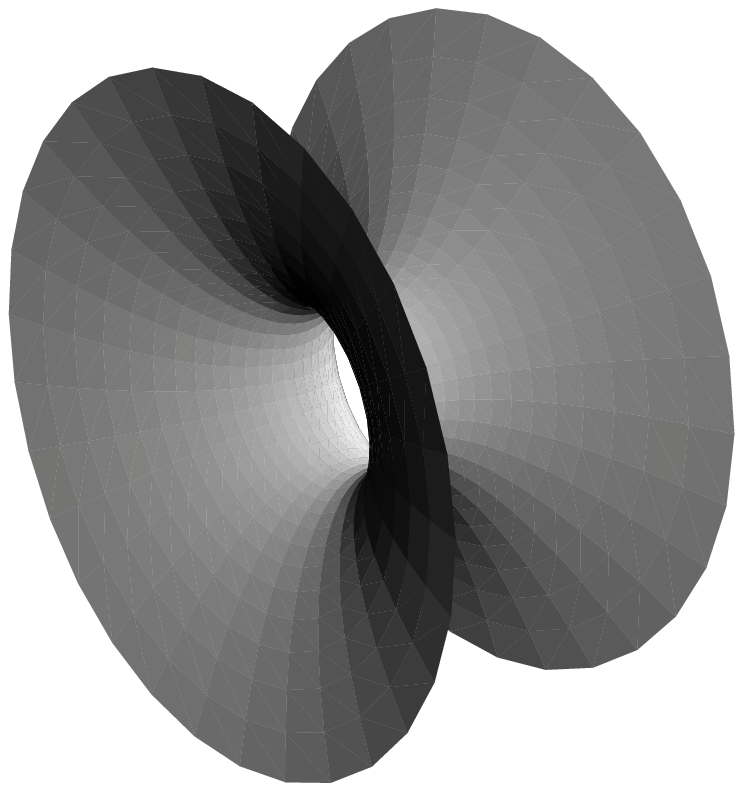}
&\includegraphics[height=5cm]{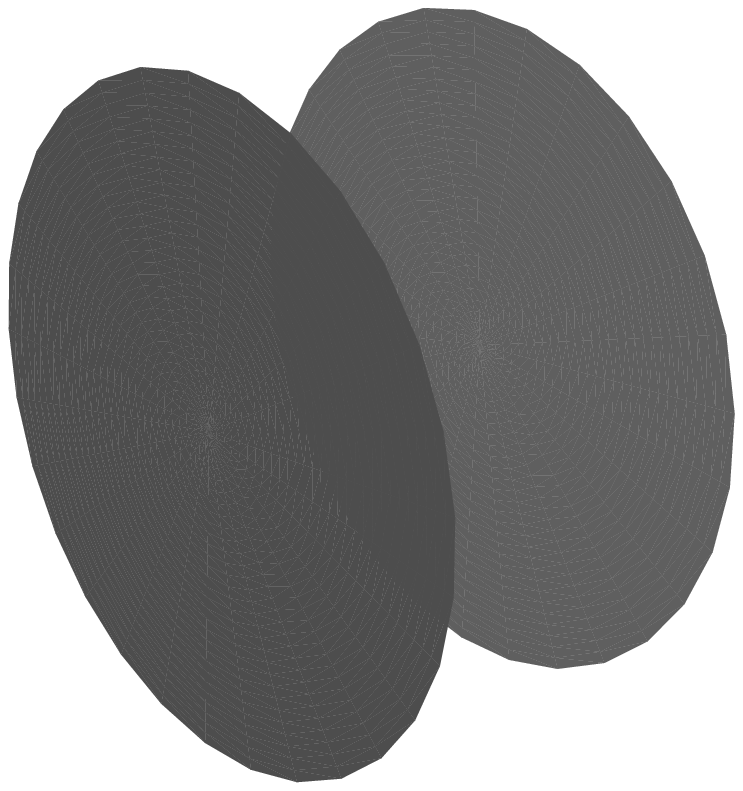}
\end{tabular}
\end{center}
\vskip -.5 cm \caption{The three equilibrium shapes of the soap
film, namely the shallow catenoid, the deep catenoid and the
Goldschmidt solution, plotted at a natural scale for $L=1$.}
\label{figa1}
\end{figure}

\begin{figure}[!t]
\begin{center}
\begin{tabular}{cc}
\includegraphics[height=5cm]{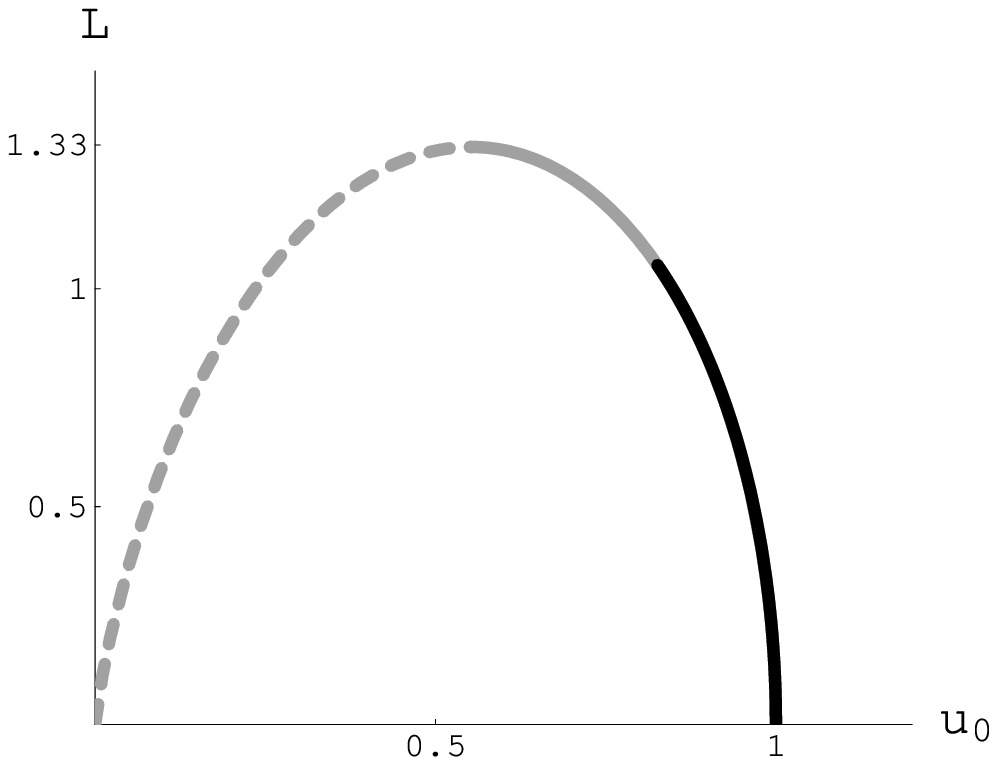}
&\includegraphics[height=5cm]{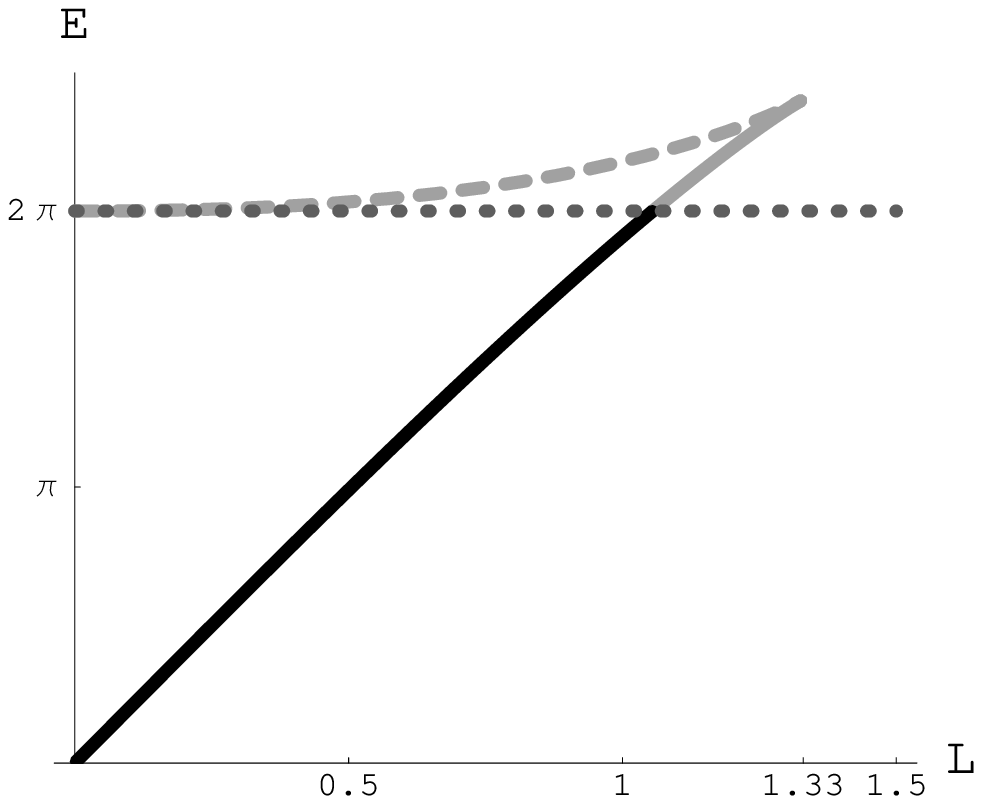}\\
(a) & (b)
\end{tabular}
\end{center}
\vskip -.5 cm \caption{Plots of $L(u_0)$ and $E(L)$, with the
solid dark, solid gray and dashed gray lines corresponding to
stable, metastable and unstable catenoid solutions. In the plot of
$E(L)$, the dotted line corresponds to the Goldschmidt solution.}
\label{figa2}
\end{figure}

\no To examine the properties of these solutions, we note that
since \eqn{a-5} gives $L(0) = L(1) = 0$ and $L^{\prime\prime}(u_0)
= - {2 \ov u_0 (1-u_0^2)^{3/2}} < 0$, $L(u_0)$ is a concave
function of $u_0$ with a single maximum for $0 \leqslant u_0
\leqslant 1$. Setting $L^{\prime}(u_0)=0$, we obtain the
transcendental equation
\ba
\label{a-7} \sqrt{1-u_0^2} \cosh^{-1} {1 \ov u_0} = 1\ ,
\ea
whose solution is $u_{0 {\rm c}} \simeq 0.552$ leading to the
maximal length $L_{\rm c} \simeq 1.325$. We must then distinguish
between two cases. For $L>L_{\rm c}$, \eqn{a-5} has no solution
for $u_0$ and the catenoid solution does not exist at all, leaving
the Goldschmidt solution as the only available one.\footnote{That
is, when the two rings supporting the soap film are stretched a
distance larger than $L_{\rm c}$ apart, the soap film will break
to form two flat circular films over each ring.} For $L < L_{\rm
c}$, on the other hand, \eqn{a-5} has {\em two} solutions for
$u_0$, the largest (smallest) of which corresponds to a
``shallow'' (``deep'') catenoid.\footnote{For example, for
$L=\ha$, these solutions have $u_0 \simeq 0.967$ (almost
cylindrical film) and $u_0 \simeq 0.076$ (highly curved film)
respectively.} Meanwhile, the Goldschmidt solution exists as well,
giving a total of three available solutions, shown in Fig.
\ref{figa1}. To examine which one is energetically favored, we
need to compare the energy of the shallow catenoid, $E(u_0)$ with
$u_0>u_{0 {\rm c}}$, with that of the Goldschmidt solution,
$E=2\pi$. The former is a decreasing function of $u_0$, becoming
equal to $2\pi$ at $u_0=\tilde{u}_{0 {\rm c}} \simeq 0.826$ where
the separation is $\tilde{L}_{\rm c} \simeq 1.055$, and hence the
lowest-energy solution is the shallow catenoid for $u_0 >
\tilde{u}_{0 {\rm c}}$ and the Goldschmidt solution for $u_0 <
\tilde{u}_{0 {\rm c}}$. To summarize, for $L>L_{\rm c}$ the only
possible solution is the Goldschmidt solution, while for $L <
L_{\rm c}$ all three solutions are available with the shallow
catenoid being favored for $L < \tilde{L}_{\rm c}$ and the
Goldschmidt solution being favored for $L > \tilde{L}_{\rm c}$.
This behavior is shown in Fig. \ref{figa2}.

\no By now, the analogy with the Wilson-loop calculations in the
main part of the paper should be obvious. Namely, the quantities
$u_0$, $L$ and $E$ correspond to the quantities denoted by the
same symbols in the Wilson-loop context, the shallow catenoid,
deep catenoid and the Goldschmidt solution correspond to the short
string, the long string and the unbound configuration respectively
and the critical values $L_{\rm c}$ and $\tilde{L}_{\rm c}$ of the
separation correspond to the maximal and screening lengths
respectively. Fig. \ref{figa2}(b) in particular is qualitatively
similar to Figs. \ref{fig1}(b) and \ref{fig4}(b) in the main part
of the paper.

\no To examine the stability of the catenoid solutions, we need to
consider small fluctuations about these equilibrium surfaces. In
\cite{durand}, the stability analysis was carried out for
perturbations normal to the surface. Here, we consider the most
general type of perturbation, which we parametrize as follows.
Writing the equilibrium surface as $F(r,z)=r-r_{\rm cl}(z)=0$,
where $r_{\rm cl}(z) = u_0 \cosh (z / u_0 )$ is the classical
solution, the unit normal to the surface is $\hat{n}={\nabla F \ov
|\nabla F|}$, while the unit tangent vectors along and
perpendicular to the azimuthal direction are $\hat{\phi}$ and
$\hat{\xi}=\hat{n} \times \hat{\phi}$, respectively. Explicitly,
we have the following transformation between the two orthonormal
frames with basis vectors $(\hat n,\hat\xi,\hat\phi)$ and $(\hat
r,\hat z,\hat \phi)$, respectively
\be
\hat{n} = {\hat{r}-r_{\rm cl}^{\prime} \hat{z} \ov \sqrt{1+r_{\rm
cl}^{\prime 2}}}\ , \qq \hat{\xi} = {\hat{z}+r_{\rm cl}^{\prime}
\hat{r}\ov \sqrt{1+r_{\rm cl}^{\prime 2}}}\ .
\ee
Expressing the most general perturbation as
\be
\d{\vec r}=\d n \hat{n} + r \d\phi \hat{\phi} + \d \xi
\hat{\xi}=\d r \hat r + r\d \phi \hat \phi + \d z \hat z\ ,
\ee
we have, in terms of our original variables,
\ba
\label{a-8}
z&=& \s_1+{\d \xi (t,\s_1,\s_2)- r^{\prime}_{\rm
cl}(\s_1) \d n (t,\s_1,\s_2) \ov \sqrt{1+r^{\prime 2}_{\rm
cl}(\s_1)}}\ ,
\nonumber\\
r&=& r_{\rm cl}(\s_1) + {\d n (t,\s_1,\s_2)
+r^{\prime}_{\rm cl}(\s_1) \d \xi (t,\s_1,\s_2) \ov \sqrt{1+r^{\prime 2}_{\rm cl}(\s_1)}}\ ,
\\
\phi &=& \s_2 + \d \phi(t,\s_1,\s_2)\ ,
\nonumber
\ea
From this form we can see that this is an $SO(2)$ transformation
with rotation angle related to the variable $\s_1$ as $\cos\th
=1/\cosh (z/u_0)$. Since this is local in space and global in time
the kinetic energy remains diagonal in this new basis.
Substituting in the original action \eqn{a-1} and expanding in
powers of the perturbations, we find that the zeroth-order term
gives the classical action, the first-order term vanishes by the
equations of motion and periodicity of $\d \phi$ in $\s_2$, and
the second-order term reads
\ba
\label{a-9} S_2 &=& - {1 \ov u_0} \int d t \int d \s_1 d \s_2
\biggl\{ {1 \ov 2} [u_0^2 (\partial_1 \d n)^2 + (\partial_2 \d
n)^2 ] \nonumber \\ && \qq\qq\qq\qq  - \: { u_0^2 \ov 2}
\cosh^2{\s_1 \ov u_0} \left(\d\dot{n}^2 + \d\dot{\xi}^2 + u_0^2
\cosh^2{\s_1 \ov u_0} \d\dot{\phi}^2\right)
 \nonumber \\ && \qq\qq\qq\qq - \: {1 \ov \cosh^2 {\s_1 \ov u_0}} \left( \d n^2 - \ha \d \xi^2 +
\sinh{\s_1 \ov u_0} \d n \d \xi \right) \nonumber \\ &&
\qq\qq\qq\qq + \: {u_0 \ov \cosh {\s_1 \ov u_0}} \left[
\partial_1 (\d n \d \xi) + \sinh {\s_1 \ov u_0} \left( \ha \partial_1 \d \xi^2 +
 \cosh {\s_1 \ov u_0} \d \xi \partial_2 \d \phi \right) \right] \nonumber \\ && \qq\qq\qq\qq +
 \: u_0^2 \cosh {\s_1 \ov u_0} ( \partial_1 \d \xi \partial_2 \d \phi
- \partial_2 \d \xi \partial_1 \d \phi) \biggr\}\ .
\ea
Although the various perturbations appear to be coupled, the
calculation of the equations of motion reveals that they actually
decouple due to an extensive cancellation of terms. For the
tangential perturbations, the equations of motion are just
\ba
\label{a-10}
\d \ddot{\xi} = \d \ddot{\phi} = 0\ .
\ea
For the normal perturbations, we rename
$(\s_1,\s_2) \to (z,\phi)$, we define $u = {z\ov u_0}$, we
separate variables according to
\be
\label{a-12}
\d n(t,u,\phi) = \Phi(u) e^{- {\rm i} \Omega t} e^{{\rm i} m \phi}\ ,
\ee
and we end up with the Sturm--Liouville equation \eqn{3-7} with
\be
p=1\ ,\qq r={2\ov \cosh^2u}-m^2\ , \qq q=\cosh^2u\ , \qq
\omega=u_0\Omega\ ,\label{a-11}
\ee
subject to the following boundary conditions at the endpoints
\ba
\label{a-14} \Phi\left(\pm {L\ov 2 u_0}\right)=\Phi \left( \pm
\cosh^{-1} {1 \ov u_0} \right) = 0\ ,
\ea
where we have used \eqn{a-5}. To investigate the stability of the
catenoid solutions, we want to determine the sign of $\omega^2$
for the lowest-energy solution to the differential equation
\eqn{3-7} with \eqn{a-11} in terms of $u_0$. Although the study of
the evolution of $\omega^2$ as a function of $u_0$ is in general a
hard task, we can obtain useful information by considering the
corresponding zero-mode problem i.e. solving \eqn{3-7} with
$\omega^2=0$. In this case, the transformation $x=\tanh u$ turns
\eqn{3-7} into an associated Legendre equation with the general
solution given by a linear combination of $P_1^m(\tanh u)$ and
$Q_1^m(\tanh u)$, $m=0,\pm 1,\pm 2,\dots$. However, the boundary
conditions \eqn{a-14} further restrict us to $m=0$, for which only
the solution proportional to $Q_1(\tanh u)$ is acceptable.
Therefore, the zero-mode solution reads
\be
\Phi_0 (u) = N Q_1(\tanh u)= N(1-u\tanh u)\ ,
\ee
where $N$ is a normalization constant. Imposing the boundary
condition \eqn{a-14} leads then to the transcendental equation
\eqn{a-7}, that is, to the condition determining the point $u_{0
{\rm c}}$ at which the length $L(u_0)$ attains its maximum!
Therefore, the equation of motion of the normal perturbations has
a zero mode if and only if $u_0$ attains the value $u_{0 {\rm c}}$
that marks the boundary between the deep and shallow catenoid
solutions. This result tremendously simplifies our problem, as it
implies that if we determine the evolution of the lowest
eigenvalue as we move infinitesimally from $u_{0 {\rm c}}$, we
will in fact have determined its sign in the whole regions to the
left and to the right of $u_{0 {\rm c}}$. Again, one may
appreciate the similarity with the case of longitudinal
perturbations in the main part of the paper.

\no To carry out this investigation, it is convenient to transform
the Sturm--Liouville problem to a Schr\"odinger one by applying the
transformation \eqn{3-8} which, for our case, reads
\ba
x=\sinh u\ ,\qq \Psi(x) =(1+x^2)^{1/4} \Phi (u)\ .
\ea
This way, we obtain the Schr\"odinger equation \eqn{3-9} where the
potential, for general values of $m$, is given by
\ba
\label{a-16} V(x) = {m^2 \ov x^2+1} - {x^2+6 \ov 4 (x^2+1)^2}\ ,
\ea
and the boundary conditions at the endpoints read
\ba
\label{a-17} \Psi ( \pm x_0 ) = 0\ ,\qq x_0 = x_0 (u_0) \equiv
{\sqrt{1-u_0^2} \ov u_0}\ .
\label{bbcc}
\ea
The potential is depicted in Fig. 8(a) and clearly does not
support any negative energy states for $|m|\geqslant 2$, which is
consistent with the discussion above. For $m=0$ it is deep enough
to support one negative energy state if its size is restricted to
a finite interval $x\in [-x_0,x_0]$, with $x<x_{0\rm c}$. For
$|m|=1$ there is no negative or zero energy state respecting the
boundary conditions as we have seen, even though the potential has
a negative part. In the new variables, the normalized zero-mode
eigenfunction occurring for $m=0$ is
\ba
\label{a-18} \Psi_0(x) ={\sqrt{6} u_{0 {\rm c}} (1-u_{0 {\rm
c}}^2)^{3/4} \ov (3 - 5 u_{0 {\rm c}}^2)^{1/2}} (1+x^2)^{1/4}
\left( 1- {x \ov \sqrt{1+x^2}} \sinh^{-1} x \right)\ .
 \ea
To determine the change in $\omega^2$ as $u_0$ deviates from $u_{0
{\rm c}}$, we may use perturbation theory. Noting that the
potential \eqn{a-16} is $u_0$--independent while the endpoints
$\pm x_0$ given in \eqn{a-17} are $u_0$--dependent, we find that
the appropriate perturbation-theory formula is
\ba
\label{a-22}
\d \omega^2 = 2 x_0^{\prime}(u_{0{\rm c}})
 \Psi_0^2(x_{0\rm c}) V(x_{0\rm c})\ \d u_0\ ,
\ea
as is read off from \eqn{gp}, appropriately adapted to a
Schr\"odinger problem for a real wavefunction in the range
$[-x_{0\rm c},x_{0\rm c}]$ and with the boundary condition in
\eqn{bbcc}. In our case $x_{0\rm c}\simeq 1.509$.
\begin{figure}[!t]
\begin{center}
\begin{tabular}{ccc}
\includegraphics[height=4.1cm]{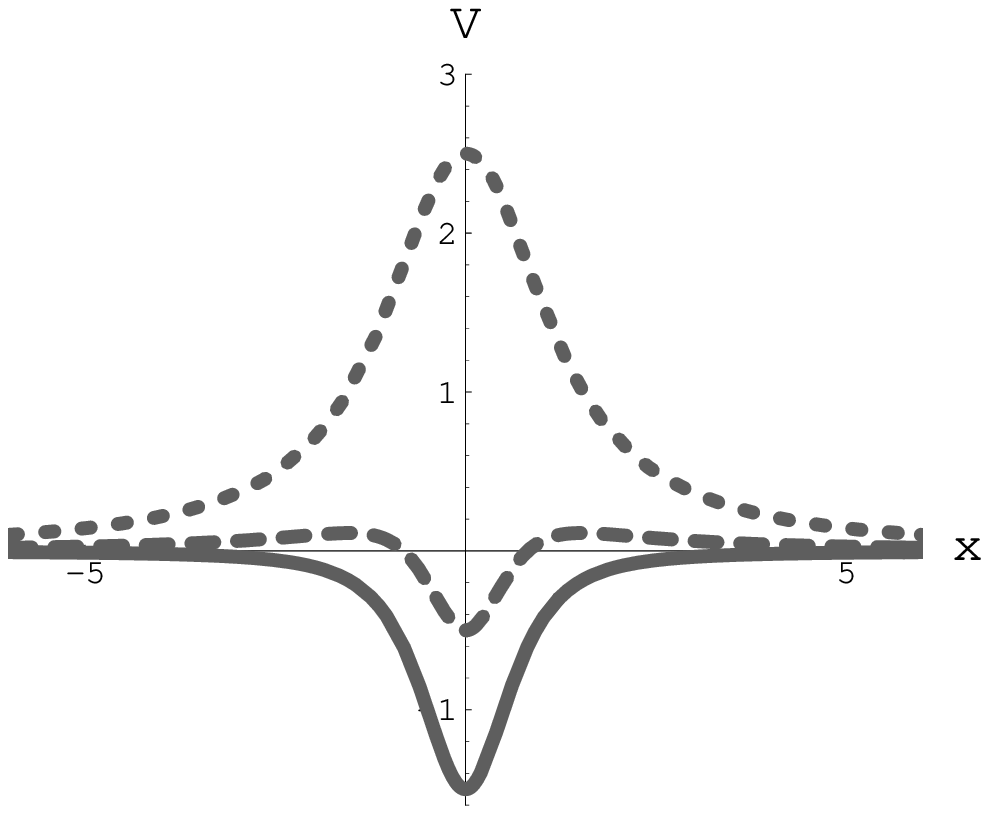} \!\!\!\!
& \!\!\!\! \includegraphics[height=4.1cm]{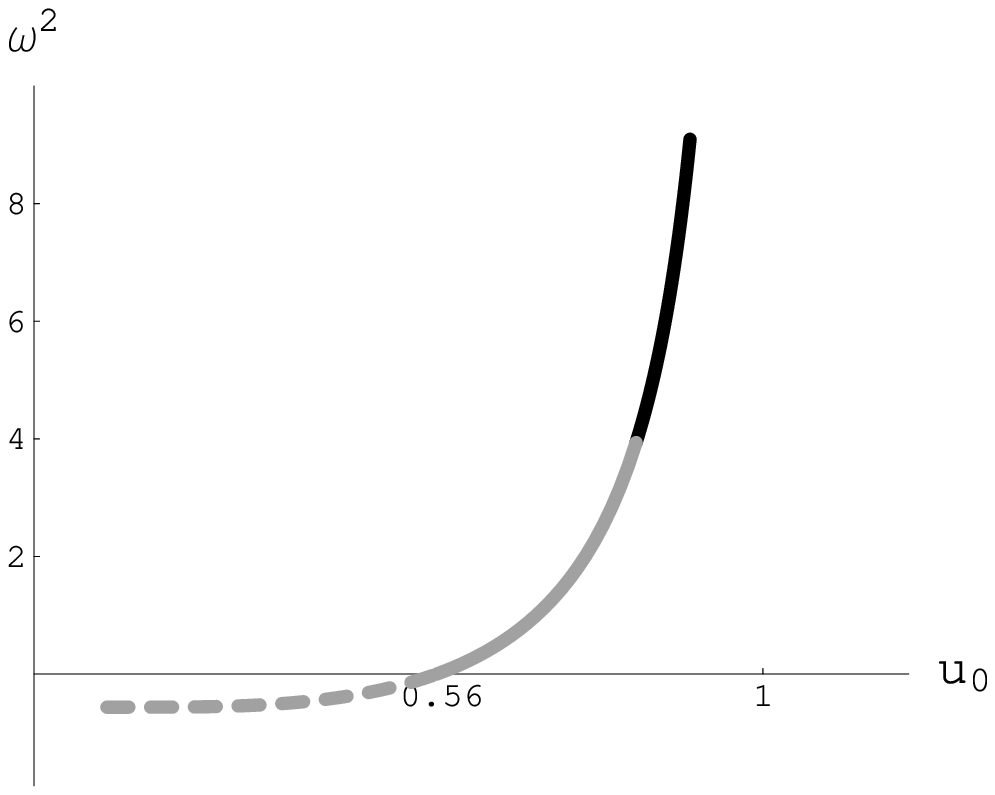} \!\!\!\!
& \!\!\!\! \includegraphics[height=4.1cm]{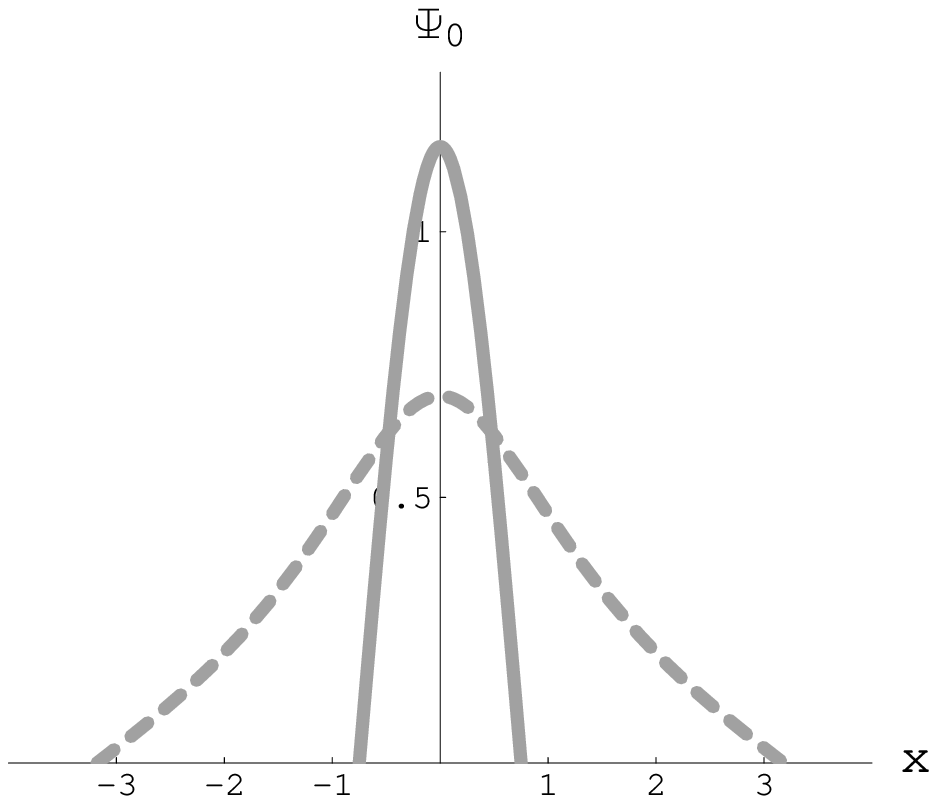} \!\!\!\! \\
\!\!\!(a)  &  (b)  &  \!\!\!\!\!(c)
\end{tabular}
\end{center}
\vskip -.5 cm \caption{(a) Schr\"odinger potentials for the normal
perturbations with angular quantum numbers $m=0$ (solid), $m=\pm
1$ (dashed) and $m=\pm 2$ (dotted). (b) Evolution of the lowest
eigenvalue $\omega^2$ with $u_0$, calculated using the shooting
method and Mathematica's \cite{mathematica} {\tt NDSolve} routine.
The slope of the curve at $u_0=u_{0 {\rm c}}$ is found to be
$4.497$, in perfect agreement with \eqn{a-23}. (c) Plots of the
normalized ground-state wavefunction $\Psi_0(x)$ for
$u_0=0.8>u_{0{\rm c}}$ (solid) and $u_0=0.3<u_{0{\rm c}}$
(dashed).} \label{figa3}
\end{figure}To calculate $\d \omega^2$, we insert \eqn{a-16} and \eqn{a-18}
into \eqn{a-22}, making repeated use of \eqn{a-7} to simplify the
resulting expressions. When the smoke clears out, we find
\be
\label{a-23} \d \omega^2 = {12 u_{0 \rm c} \ov 3-5 u_{0 \rm c}^2}
\d u_0 \simeq 4.497\ \d u_0\ .
\ee
Hence, as we move to the right (left) of $u_{0{\rm c}}$, the
eigenvalue $\omega^2$ becomes positive (negative) and therefore
the shallow catenoid is stable while the deep catenoid is
unstable, as expected. This behavior is confirmed by a numerical
analysis, shown in Fig. \ref{figa3}.

\end{document}